# Non-unitary Quantum Physical Unclonable Functions: Modelling, Simulation, and Evaluation under Open Quantum Dynamics

Mohammadreza Vali[a], Hossein Aghababa[a,b], Nasser Yazdani[a]

[a]*School of Electrical and Computer Engineering, College of Engineering, University of Tehran, Tehran, 1439957131, Iran*
[b]*Department of Engineering, Loyola University Maryland, Baltimore, MD 21210, USA*

**Abstract**

Physical Unclonable Functions (PUFs) have long served as a cornerstone for hardware-based security, leveraging intrinsic manufacturing randomness to generate device-unique responses for authentication and key generation. However, as machine learning and side-channel attacks increasingly undermine classical PUF assumptions, new approaches are required to achieve fundamental unforgeability. Quantum mechanics offers a natural foundation for this goal through its inherent randomness and the no-cloning theorem, motivating the development of Quantum Physical Unclonable Functions (QPUFs). Yet, most existing QPUF models rely on idealized unitary dynamics, neglecting the unavoidable non-unitary interactions that occur in real quantum hardware due to decoherence and dissipation.

This study introduces and analyzes a new class of non-unitary Quantum Physical Unclonable Functions (QPUFs), which exploit open quantum system dynamics as the core mechanism of security. Three architectures are proposed and evaluated: the Dissipative QPUF (D-QPUF), which harnesses amplitude damping as an entropy source; the Measurement-Feedback QPUF (MF-QPUF), which integrates mid-circuit measurements and conditional unitaries to produce stochastic evolution; and the Lindbladian QPUF (L-QPUF), which generalizes the concept using the Lindblad master equation and Trotter–Suzuki decomposition to model Markovian noise-driven evolution.

Simulation results demonstrate that these non-unitary designs maintain high levels of uniqueness, uniformity, and unforgeability, while exhibiting controlled trade-offs in reliability due to stochastic channel effects. Importantly, the L-QPUF architecture achieves exponential modelling resistance under limited challenge–response access, indicating that non-unitary evolution can serve as a practical foundation for post-quantum hardware authentication. By reframing environmental noise as a constructive security resource rather than a limitation, this work establishes a theoretical and computational framework for noise-aware quantum hardware authentication. The findings highlight that the integration of non-unitary dynamics not only reflects physical reality but also opens a promising pathway toward scalable, inherently unclonable quantum devices.

## 1. Introduction

The growing demand for secure hardware authentication, tamper resistance, and device identification has driven decades of research in Physical Unclonable Functions (PUFs)—hardware primitives that exploit intrinsic manufacturing randomness to generate unique and irreproducible responses [1, 2]. These responses, derived from complex physical processes rather than stored digital keys, form the basis of lightweight authentication and key generation systems for embedded and Internet-of-Things (IoT) devices [3, 4, 5, 6]. However, classical PUFs, while successful in many applications, face fundamental limitations in scalability, attack resistance, and long-term reliability. The increasing sophistication of machine learning–based modelling and side-channel attacks has made it increasingly difficult to ensure that classical PUFs remain unclonable under realistic adversarial assumptions [7, 8].

In parallel, advances in quantum technologies have opened a new frontier for both hardware security and quantum cryptography [9, 10, 11]. Quantum systems inherently exhibit properties—such as measurement randomness, superposition, and the no-cloning theorem—that naturally align with the concept of unforgeability [12, 13]. These features inspired the development of Quantum Physical Unclonable Functions (QPUFs), which extend the PUF concept into the quantum domain [14, 15]. In contrast to their classical counterparts, QPUFs leverage quantum mechanical irreversibility and state collapse as entropy sources, offering fundamentally stronger protection against duplication and emulation [16, 17, 18]. Recent experimental progress in quantum information processing has also demonstrated the practical realization of noisy, entangled systems on IBM quantum hardware, providing an empirical foundation for evaluating quantum unforgeability under realistic decoherence [40].

Despite their promise, most QPUF designs to date have been modelled using closed quantum systems governed by unitary evolution, assuming idealized conditions free from decoherence or dissipation [17, 19, 20]. This assumption simplifies analysis but overlooks the fact that real quantum devices are inherently open systems, continually interacting with their environment. Such interactions introduce non-unitary dynamics, manifesting as noise, dephasing, or relaxation processes [13, 21]. Traditionally, these effects have been viewed as detrimental to quantum computation and communication, as they degrade fidelity and coherence [10, 22]. On the other hand, quantum noise can also be exploited constructively to enhance the robustness of information sharing and secure communication across multi-party networks, suggesting its potential role as an enabler of unforgeability in noisy quantum systems [42] and other work argues that environmental noise can be reinterpreted as a source of intrinsic physical entropy—a property that can be harnessed for security rather than mitigated [23].

This perspective forms the foundation of the present study, which introduces and analyzes a new class of non-unitary Quantum Physical Unclonable Functions (QPUFs). Instead of suppressing decoherence, these architectures leverage it as the fundamental mechanism for unforgeability. The research systematically explores three distinct designs:

- **Dissipative QPUF (D-QPUF):** A non-unitary model that explicitly incorporates amplitude-damping channels to exploit natural quantum decay as an entropy source [24, 25].
- **Measurement-Feedback QPUF (MF-QPUF):** A hybrid model that integrates mid-circuit measurements and classical feedback to generate stochastic, history-dependent evolution paths [13, 26].
- **Lindblad QPUF (L-QPUF):** A generalized framework based on open quantum system theory, modelling device behaviour through Lindblad master equations and their channel decompositions [24, 25, 27].

Each design progressively extends the theoretical and practical boundaries of QPUF implementation, moving from simple noise-driven processes toward structured open-system dynamics that can be realized on near-term quantum hardware [10, 11]. By embedding security directly into the non-unitary evolution of quantum states, these designs propose a noise-aware, hardware-rooted approach to quantum authentication—one that aligns with the realities of today's imperfect quantum processors.

To evaluate their feasibility, the study employs density matrix simulations and quantum channel reconstruction techniques to assess key PUF metrics, including uniqueness, uniformity, and reliability [13, 22]. The results demonstrate that controlled non-unitarity not only preserves but can enhance unforgeability, offering exponential resistance to modelling attacks under limited challenge–response access [18, 19, 28].

The main contributions of this research are threefold:

- **Theoretical Contribution:** Formalization of non-unitary QPUFs grounded in open quantum system dynamics, extending the PUF paradigm beyond unitary evolution [14, 25, 29].
- **Design Contribution:** Introduction of three distinct architectures—D-QPUF, MF-QPUF, and L-QPUF—each demonstrating a unique method of embedding physical irreversibility into the challenge–response process [2, 24, 26].
- **Analytical Contribution:** Comprehensive simulation-based evaluation of the proposed models, revealing practical trade-offs between entropy generation, reproducibility, and hardware feasibility [22, 24, 27].

The remainder of this paper is organized as follows. Section 2 provides important definitions and concepts from quantum computing, open quantum systems and PUFs. Section 3 presents a detailed literature review of classical and quantum PUF research, including recent advances in open-system simulation. Section 4 introduces the proposed non-unitary QPUF architectures, followed by Section 5, which discusses the simulation setup and results. Section 6 concludes the study by summarizing the implications of non-unitary QPUFs for secure hardware design, and Section 7 outlines the key challenges and directions for future work.

Through this exploration, the paper aims to establish non-unitary QPUFs as a viable, next-generation approach to secure device authentication—one that redefines noise and dissipation not as threats to quantum information, but as core enablers of physical unforgeability in the quantum era.

## 2. Preliminaries

This section introduces the basic concepts and terminology used throughout this dissertation, focusing on quantum computing principles and the theoretical foundations of Physical Unclonable Functions (PUFs) and Quantum PUFs (QPUFs). The descriptions and notations are primarily based on Nielsen and Chuang book [13] and Preskill lecture notes [30].

### *2.1. Basic definitions*

In this sub-section, we are going to provide some definitions and concepts from quantum computing and open quantum systems which can be highly beneficial in order to fully comprehend the rest of this paper.

- **Qubit and Quantum states:** A qubit is the fundamental unit of quantum information. Unlike a classical bit, which can be either 0 or 1, a qubit can exist in a superposition of both basis states, expressed as $|\psi\rangle = \alpha|0\rangle + \beta|1\rangle$, where $\alpha, \beta \in \mathbb{C}$ and satisfy the normalization condition $|\alpha|^2 + |\beta|^2 = 1$, ensuring total probability equals one.

- **Hilbert space:** All quantum states exist in a complex vector space known as a Hilbert space. Each valid quantum state corresponds to a unit vector in this space, equipped with an inner product that defines the geometry of quantum state evolution.

- **Trace of Quantum states:** The trace of a matrix is the sum of its diagonal elements. In quantum mechanics, it is used to compute probabilities, expectation values, and to describe mixed-state evolution.

- **Pure and Mixed states:** A pure state is fully characterized by a single state vector in Hilbert space, with purity condition $\mathrm{Tr}(\rho^2) = 1$. On the other hand, A mixed state, by contrast, represents a probabilistic ensemble of pure states, given by $\rho = \sum_i p_i |\psi_i\rangle\langle\psi_i|$ where $\sum_i p_i = 1$ and satisfies $\mathrm{Tr}(\rho^2) < 1$.

- **Uhlmann fidelity and Trace distance:** The Uhlmann fidelity between two quantum states $\rho$ and $\sigma$ quantifies their similarity

$$F(\rho, \sigma) = \left(\mathrm{Tr}\sqrt{\sqrt{\rho}\sigma\sqrt{\rho}}\right)^2 \qquad (1)$$

and their distinguishability can also be measured via the trace distance

$$D(\rho, \sigma) = \frac{1}{2} \parallel \rho - \sigma \parallel_1 \qquad (2)$$

- **Diamond distance:** The diamond norm quantifies the distance between two quantum channels, considering auxiliary qubits. It is a standard measure for comparing real noisy channels to their ideal unitary counterparts.

- **Hermitian Operators and Hamiltonians:** an operator $H$ is Hermitian if $H = H^\dagger$, implying real eigenvalues and physical observability. In quantum mechanics, the Hamiltonian is a Hermitian operator that represents the total energy of the system. Its dynamics obey the time-dependent Schrödinger equation:

$$i\hbar \frac{d}{dt} \mid \psi(t)\rangle = H \mid \psi(t)\rangle \qquad (3)$$

- **Unitary and Non-Unitary Operations:** Closed-system evolution is described by a unitary operator $U$ satisfying $U^\dagger U = I$. Non-unitary operations do not satisfy this condition and model open-system or noisy dynamics that are irreversible.

- **Measurement and Projective Measurement:** Quantum measurement collapses a state into one of the basis states, producing classical outcomes. A projective measurement is represented by operators $\{M_m\}$ satisfying

$$\sum_m M_m^\dagger M_m = I. \qquad (4)$$

The probability of outcome $m$ is $p(m) = \langle\psi \mid M_m^\dagger M_m \mid \psi\rangle$, and the post-measurement state is

$$\frac{M_m \mid \psi\rangle}{\sqrt{p(m)}}. \qquad (5)$$

- **Density Matrix Formalism:** The density matrix provides a unified representation for both pure and mixed states. It satisfies $\rho^\dagger = \rho$, $\rho \geq 0$, and $\mathrm{Tr}(\rho) = 1$.

- **Open Quantum Systems and the Lindblad Equation:** An open system interacts with its environment, leading to non-unitary evolution described by the Lindblad (or GKSL) master equation:

$$\frac{d\rho}{dt} = -i[H,\rho] + \sum_k \left(L_k \rho L_k^\dagger - \frac{1}{2}\{L_k^\dagger L_k, \rho\}\right), \qquad (6)$$

where $L_k$ are Lindblad (jump) operators representing environmental interactions such as relaxation or dephasing.

- **Decay rate and Relaxation time:** The decay rate characterizes how rapidly coherence or energy dissipates. The relaxation time $T_1$ measures population decay, while the phase coherence time $T_2$ quantifies phase stability, related via equation 7, where $T_\phi$ is the pure phasing time.

$$\frac{1}{T_2} = \frac{1}{2T_1} + \frac{1}{T_\phi} \qquad (7)$$

- **Quantum Channels and Kraus Representation:** A quantum channel is a CPTP map of the form:

$$\mathcal{E}(\rho) = \sum_k E_k \rho E_k^\dagger \text{ ,with } \sum_k E_k^\dagger E_k = I. \qquad (8)$$

Here, $E_k$ are the Kraus operators describing stochastic noise processes.

- **Quantum noise models:** in this study we have investigated three noise models including Amplitude damping with Kraus operators

$$K_0 = \begin{bmatrix} 1 & 0 \\ 0 & \sqrt{1-p} \end{bmatrix} \qquad (9)$$

and

$$K_1 = \begin{bmatrix} 0 & \sqrt{p} \\ 0 & 0 \end{bmatrix}, \qquad (10)$$

Phase damping with Kraus operators

$$K_0 = \sqrt{1-q}I \qquad (11)$$

and

$$K_1 = \sqrt{q}\sigma_z, \qquad (12)$$

and Depolarizing with Kraus operators

$$K_0 = \sqrt{1-p}I \qquad (13)$$

and

$$K_{1,2,3} = \sqrt{\frac{p}{3}}\{X, Y, Z\}. \qquad (14)$$

- **Markovian Systems:** Markovian systems are memoryless, with evolution dependent only on their current state.

- **NISQ devices:** NISQ (Noisy Intermediate-Scale Quantum) devices represent current quantum processors with limited qubits and significant noise.

- **Quantum State Tomography:** A process for reconstructing the density matrix of a quantum state by performing measurements in multiple bases and statistically inferring the underlying state.

## *2.2. Formal definition of a Physical Unclonable Function*

A Physical Unclonable Function (PUF) is a physical mapping $C \to R$, transforming an input challenge $C$ into a response $R$, where the transformation depends on intrinsic, uncontrollable physical variations of the device [30]. These properties are infeasible to reproduce, even by the original manufacturer.

The essential features of an ideal PUF are:

- **Unforgeability:** No physical copy can reproduce identical behavior.
- **Unpredictability:** Given many challenge–response pairs, predicting unseen responses is computationally infeasible.
- **Stability:** Responses remain consistent across environmental variations.

## *2.3. Performance metrics*

Beside randomness, Following [8, 14] three standard metrics are used to evaluate PUFs and QPUFs:

- **Uniformity:** Measures the balance of output responses. For a device $Q$ and input challenge $\rho$ producing output $\rho_{\text{out}} = Q(\rho)$, uniformity is:

  $$\mathbb{E}_\rho \left[ \| Q(\rho) - \frac{I}{d} \|_\diamond \right], \qquad (15)$$

  where the ideal value is 0. Alternatively, it may be expressed using fidelity $F(\rho_{\text{out}}, I/d) = 1$. In classical systems, this corresponds to an equal distribution of 0s and 1s (ideally 50%).

- **Uniqueness:** Quantifies how distinguishable two devices $Q_i$ and $Q_j$ are for identical challenges:

  $$\mathbb{E}_\rho [\| Q_i(\rho) - Q_j(\rho) \|_\diamond], \qquad (16)$$

  with the ideal value equal to 2. In fidelity form, $F(\rho^i_{\text{out}}, \rho^j_{\text{out}}) = 0$. Classically, this corresponds to an average Hamming distance near 50%.

- **Reliability:** Measures reproducibility under repeated challenges. For outputs $\rho_{\text{out}}$ and $\sigma_{\text{out}}$ under different noise conditions:

$$\text{Reliability} = 1 - \mathbb{E}_{\rho}[D(\rho_{\text{out}}, \sigma_{\text{out}})], \qquad (17)$$

where $D$ denotes trace or Hamming distance. The ideal value corresponds to $D = 0$ (or $F = 1$). In classical systems, this equates to 100% bit-wise reproducibility.

## 3. Literature review

The concept of the Physical Unclonable Function (PUF) was first introduced by Gassend et al. [1], who proposed exploiting random variations inherent in semiconductor manufacturing as unique hardware "fingerprints." Their design demonstrated that delay variations in integrated circuits could be used for device authentication, establishing the foundation of hardware-based physical security primitives. Subsequent studies explored key extraction from such intrinsic variability. Lim et al. [31] examined secure key derivation from on-chip randomness, while Devadas et al. [5] formalized the use of PUFs for device authentication and secret generation, proposing an evaluation framework for PUF reliability and security.

With the growing adoption of PUFs in embedded systems and IoT devices, researchers sought efficient implementations tailored to constrained environments. Lightweight FPGA-based architectures, including SR-Latch-based PUFs [4] and LFSR-based designs [3], were introduced to achieve favorable trade-offs among area, power, and reliability. These advancements were driven by challenges such as ensuring stability under environmental variation, improving resistance against modeling and side-channel attacks, and reducing implementation overhead for large-scale deployment.

Despite significant progress, classical PUFs have increasingly struggled against sophisticated adversarial models. Advances in machine learning–based attacks [7] enabled high-accuracy prediction of PUF responses, compromising their assumed unforgeability. As classical designs rely on deterministic physical effects, they face an asymptotic vulnerability—given sufficient data, an attacker can approximate their challenge–response mapping. This limitation has motivated the search for security primitives rooted in more fundamental, non-deterministic physical laws.

The advent of quantum technologies introduced a transformative paradigm for hardware authentication, leading to the emergence of Quantum Physical Unclonable Functions (QPUFs). Beyond cryptography, this field extends to protecting quantum hardware from tampering and spoofing [11, 32]. The first explicit QPUF model, the Quantum Readout PUF (QR-PUF), was proposed by Skoric [15], introducing a protocol in which quantum challenges yield classical responses through measurement. This approach provided resistance to replay-based attacks, though its practical realization was limited by the requirement of quantum memory.

Subsequent work refined QPUF architectures to overcome such limitations. Designs based on $R_y$ and Hadamard gates [2], as well as stabilizer-based QPUFs, were proposed to enhance response reproducibility. Other studies developed classical–quantum hybrid authentication protocols [28], eliminating the need for quantum memory. Hybrid identification frameworks for quantum networks [34] and QPUF-based identification protocols [16] further extended QPUFs toward practical networked applications.

Interest in integrating QPUFs with IoT and cloud computing systems has also expanded rapidly. QPUFs have been explored for memory verification [35], as demonstrated in the Soteria protocol [36], which mitigates replay and proxy attacks. Semiconductor-based implementations, particularly resonant tunneling diodes (RTDs) [37], have been investigated for low-cost, reproducible QPUF behavior. Other designs combined $X$, $CX$, $R_y$, and $H$ gates with Quantum Key Distribution (QKD) [6, 38] for securing smart grids and industrial IoT systems. Additionally, Hadamard- and decoherence-based QPUFs [20] have been explored for cloud authentication, where tunable gate rotation angles enhance uniqueness and entropy.

A major theoretical milestone was achieved by Doosti et al. [14, 29], who provided formal security definitions for QPUFs by adapting the classical notions of existential, selective, and exponential unforgeability. They introduced an attack framework based on the Universal Quantum Emulator algorithm, proving that no *unitary-only* QPUF can achieve existential unforgeability, though selective unforgeability remains attainable. This insight identified the development of non-unitary QPUFs as a key open challenge. Subsequent theoretical work introduced non-unitary models based on random von Neumann measurements [18], proving that such systems can achieve existential unforgeability due to their inherent irreversibility.

Alternative designs considered Haar-random unitary ensembles [17]—such as T-structure QPUFs constructed from X, Y, and CZ gates—to strengthen modeling resistance. Analyses showed that non-fixed, randomized QPUFs [14] require exponentially larger challenge–response sets for successful modeling, as adversarial information gain remains fundamentally limited. In parallel, studies on controlled quantum broadcasting and cluster-state generation [41] have revealed how structured multi-qubit entanglement can amplify entropy and resilience under noise, suggesting similar benefits for non-unitary QPUF architectures. Despite these advances, most QPUF studies have been restricted to closed, unitary dynamics, neglecting the unavoidable impact of environmental decoherence and noise. As summarized in Table 1, previous PUF and QPUF architectures exhibit diverse trade-offs in uniformity, uniqueness, and reliability, depending on their underlying gate structures and qubit counts. These results provide a quantitative baseline for evaluating our proposed non-unitary QPUF architectures.

***Table 1.*** *Results of uniformity, uniqueness, and reliability metrics along with the number of qubits and type of gates used in previous QPUF and PUF designs.*

| Design | Classical/Quantum | #Qubits | Gates | Uniformity | Uniqueness | Reliability |
|---|---|---|---|---|---|---|
| 8 | Quantum | 4 | $R_y$, H, CX | - | 53.33 | 99 |
| 27 | | 3 | $R_y$, H, I | - | 55 | 96 |
| 22 | | 4 | $R_y$, H | - | 25 | 40 |
| 19 | | 8 | $R_x, R_y, R_z$ | - | 30 | 94.4 |
| 21 | | 8 | $R_y$, H, CX | - | 50 | 87 |
| 30 | | - | no gates, uses RTD | - | 49 | 93 |
| 12 | Classical | - | Inverters, pass transistors, 2:1 multiplexers, race arbiter latch | 51 | 38 | 90 |
| 18 | | - | 2:1Chains of multiplexers, D-type flip-flop arbiter latch | 50.12 | 46 | 99 |
| 2 | | - | Multiplexers, XOR gates, D flip-flops | 50.3 | 51 | 93 |
| 3 | | - | NOR-based SR latches, selector multiplexers | 49.9 | 49 | 80 |

Recent progress in open quantum system simulation has enabled realistic modeling of non-unitary evolution. Techniques involving ancillary qubits and partial tracing [24] allow the encoding of environmental effects into effective non-unitary channels. Such approaches directly support the implementation of Dissipative (D-QPUF) and Lindbladian (L-QPUF) architectures on near-term quantum devices. Advanced methods for simulating dissipative systems—such as variational algorithms for steady-state preparation [27] and Trotter–Suzuki decomposition of Lindblad evolution [25]—have further expanded the practical feasibility of these models.

Understanding and characterizing quantum noise has become central to this paradigm shift. On real devices, decoherence and relaxation dominate circuit behavior, limiting fidelity and reproducibility [21]. Quantum process tomography [22] enables reconstruction of device-specific noise channels, revealing stable, reproducible characteristics that can act as physical fingerprints. Rather than being treated purely as detrimental, such noise can serve as an intrinsic entropy source. Controlled noise injection has been studied not only for error mitigation but also as a security mechanism [9, 23], protecting hardware against modeling and reverse-engineering attacks while improving generalization in variational algorithms.

This emerging perspective reinterprets noise and dissipation as computational resources for unforgeability. The D-QPUF proposed in this work exploits intrinsic amplitude damping as a randomness source, while the L-QPUF formalizes this idea through the Lindblad master equation, achieving stability and exponential unforgeability. Together with the Measurement-Feedback QPUF (MF-QPUF), which introduces classical feedback into the quantum evolution, these architectures embody the transition from deterministic, unitary QPUFs to stochastic, physically grounded non-unitary systems.

In summary, the literature reveals a clear evolution—from early classical PUFs exploiting process variations, through unitary QPUFs emphasizing reversibility, to non-unitary, open-system QPUFs that explicitly integrate environmental interactions. The convergence of open quantum system theory, quantum noise characterization, and quantum hardware security has thus established the foundation for this study, which advances the field by exploiting noise-driven dynamics as a novel mechanism for quantum physical unforgeability.

## 4. Non-Unitary Quantum Physical Unclonable Functions

In designing Quantum Physical Unclonable Functions (QPUFs), one of the most fundamental considerations is whether the internal circuit should be modelled as a purely unitary evolution—representing an ideal closed quantum system—or as a non-unitary evolution that incorporates the effects of environmental interaction and noise. Unitary circuits, while mathematically elegant, are entirely reversible, as each transformation is governed by a unitary matrix $U$ satisfying $U^{\dagger}U = I$. The mapping $|\psi\rangle \mapsto U|\psi\rangle$ preserves both probability and coherence, and thus the overall transformation can, in principle, be inverted or learned. This reversibility, however, introduces a security vulnerability: the predictability of unitary circuits makes them more susceptible to modelling and machine-learning-based attacks, since an adversary can efficiently approximate the system behaviour using a limited number of challenge-response pairs.

In contrast, a non-unitary circuit is characterized by a completely positive trace-preserving (CPTP) channel of the form equation 8. Such mappings are intrinsically irreversible, as information about the system's prior state is partially lost into the environment. This irreversibility, rather than being a drawback, becomes a valuable cryptographic resource: it strengthens unforgeability and makes the channel much harder to reconstruct or invert.

In a non-unitary QPUF, irreversibility is therefore not an undesired side effect but a designed feature, deliberately employed to enhance security.

Non-unitarity can emerge in several distinct physical situations. It arises when intermediate measurements are performed within the circuit, since measurement collapses quantum superpositions and destroys entanglement. It also appears naturally as a consequence of noise and decoherence, where the system interacts with its surrounding environment—such as through amplitude damping, dephasing, or thermal relaxation—none of which can be described by a single unitary transformation. Another common origin of non-unitarity is the use of ancillary qubits that are later discarded. When the system's state is obtained by tracing out ancillas, as in $\rho_{\text{sys}} = \text{Tr}_{\text{anc}}[U(\rho \otimes | 0\rangle\langle 0 |)U^{\dagger}]$, the resulting subsystem dynamics become non-unitary even though the total joint evolution remains unitary. Non-unitarity can also be introduced intentionally through probabilistic gate application—for example, applying a Pauli-$X$ gate with probability $p$ and omitting it otherwise leads to a channel $\rho \mapsto (1-p)\rho + pX\rho X$, which cannot be represented by a single unitary operator. In all such cases, the circuit evolution cannot be described by an invertible operator and is best modelled within the CPTP framework. This representation embeds irreversibility and information erasure at a fundamental level, making it ideally suited for QPUF applications where unpredictability and unforgeability are desired properties.

The distinction between unitary and non-unitary transformations can be analysed more precisely by considering their respective parameter spaces. In a Hilbert space of dimension $d$, unitary channel is defined by an operator $U \in U(d)$ and is determined—up to a global phase—by approximately $d^2 - 1$ real parameters, corresponding to the dimension of the special unitary group $SU(d)$. A general CPTP channel, on the other hand, can be represented by its Choi matrix $J_{\Lambda} \in \mathbb{C}^{d^2 \times d^2}$, which is positive semidefinite and satisfies $\text{Tr}_{\text{out}} J_{\Lambda} = I_{\text{in}}$. such a matrix typically involves on the order of $d^4 - d^2$ real parameters. For a single-qubit system, the difference is particularly striking: a unitary operation requires only three parameters (e.g., Euler angles on the Bloch sphere), whereas a general quantum channel requires twelve parameters, which can be expressed as an affine transformation $\vec{r} \mapsto M\vec{r} + \vec{t}$, with a $3 \times 3$ matrix $M$ and a three-dimensional translation vector $\vec{t}$. This exponential growth in the number of parameters has significant security implications.

In process tomography, reconstructing a channel to accuracy $\epsilon$ requires a number of experiments that scales roughly as $N \gtrsim C \cdot P/\epsilon^2$, where $P$ is the number of independent parameters and $C$ depends on the estimation method. For unitary processes $P \sim d^2$, while for general CPTP channels $P \sim d^4$. Thus, the number of samples needed to reconstruct a non-unitary channel grows exponentially with the number of qubits, making full characterization computationally intractable. From a security perspective, this implies that an adversary attempting to learn or clone a non-unitary QPUF would require exponentially more challenge–response queries than in the unitary case, leading to substantially higher resilience against machine-learning or tomography-based attacks.

A complementary perspective arises from the notion of channel distinguishability. For two unitary channels $U$ and $V$, their distance can be expressed using the diamond norm $\| U - V \|_{\diamond} = 2\sqrt{1 - | \text{Tr}(U^{\dagger}V) |^2/d^2}$. When environmental noise is introduced, this distinguishability decreases, making the channels harder to tell apart. Although this complicates verification by the legitimate verifier, it also improves security since each query yields less information to the adversary. Hence, non-unitarity naturally enhances unforgeability by limiting the information leakage per interaction. Nonetheless, care must be taken to balance non-unitary effects, as excessive or challenge-independent noise can reduce the verifier's ability to reliably authenticate legitimate responses. Ideally, the degree of non-unitarity should depend on the specific challenge or device instance, preserving both unpredictability and repeatability.

Formally, a non-unitary QPUF can be described as a quantum channel E\mathcal{E}E acting on the challenge state $\rho_{\text{in}}$, such that equation 8 holds true. The output is a statistical mixture over several physical trajectories, making the mapping inherently irreversible. Alternatively, when the evolution occurs continuously in time, the same dynamics can be represented through the Lindblad master equation (equation 6), where $H$ denotes the Hamiltonian, $L_j$ the jump operators, and $\gamma_j$ the associated rates. The resulting semigroup $e^{Lt}$ defines a CPTP map equivalent to a non-unitary channel, showing that Kraus and Lindblad representations are two complementary formulations of the same open-system evolution.

At the circuit level, the implementation of a non-unitary QPUF involves several stages. The process begins with the preparation of the challenge quantum state $\rho_{\text{in}}$, followed by the application of one or more hidden unitary layers intrinsic to the hardware. These are interleaved with non-unitary operations, realized either as explicit Kraus maps or as dissipative evolutions governed by the Lindblad equation. Intermediate measurements may be inserted to introduce stochastic feedback, conditionally determining subsequent gate operations. Finally, the device performs a measurement and generates the response, which can be a mixed quantum state or a classical bitstring recorded in a challenge–response table. Unlike unitary QPUFs, where the same challenge always produces the same deterministic output, the responses of non-unitary QPUFs are influenced by environmental factors such as temperature, device imperfections, and temporal fluctuations. This introduces additional entropy that makes responses less predictable and substantially more difficult to forge, even for adversaries with partial knowledge of the circuit.

However, this stochasticity also introduces a trade-off between security and reliability. If noise levels fluctuate excessively or remain independent of the input challenge, they can degrade authentication accuracy or enable statistical modelling of the system's behaviour. Hence, practical non-unitary QPUFs must be carefully designed so that non-unitary effects are both challenge-dependent and hardware-specific, maintaining repeatable yet unpredictable responses over time.

The theoretical framework developed above establishes non-unitarity as a constructive principle rather than a limitation. By harnessing irreversibility, environmental coupling, and probabilistic evolution, non-unitary QPUFs achieve higher resistance against cloning and modelling attacks compared to their unitary counterparts. The following sections introduce three specific realizations of this concept—D-QPUF, MF-QPUF, and L-QPUF—each employing a distinct mechanism for introducing non-unitary dynamics through decoherence, measurement feedback, or Lindblad-type dissipative evolution, respectively. Together, these models demonstrate that non-unitary quantum channels provide a feasible and powerful foundation for next-generation physical unclonable functions.

### *4.1. Dissipative channel Quantum Physical Unclonable Function*

The Dissipative channel Quantum Physical Unclonable Function (D-QPUF) constitutes the first proposed design in this study, where the evaluation function is realized through a non-unitary, dissipative quantum channel. Each device is characterized by a fixed yet secret completely positive trace-preserving (CPTP) map, denoted as $\Lambda_{\text{id}}$, that models intrinsic quantum noise. This map acts as a combination of local single-qubit noise processes such as decoherence, dephasing, and depolarization, providing a realistic representation of physical device imperfections. When the channel acts on an input challenge state $\rho_{\text{in}}$, the measurement of the resulting quantum state in the computational basis yields the device's unique classical response.

The D-QPUF design assumes multiplicative local noise parameters for each qubit, an approach that is both experimentally plausible and analytically tractable. This parameterization enables the derivation of learnability bounds and quantifiable security measures. Formally, the device generation process, denoted as $QGen$, samples noise parameters for each qubit $j \in 1, \dots, n$, defining the decoherence probability $p_j$, dephasing probability $q_j$, and depolarization rate $\lambda_j$. The overall channel can be expressed as a tensor product of local noisy maps,

$$\Lambda_{\text{id}}(\rho) = \left( \bigotimes_{j=1}^{n} D_{p_j}^{\text{amp}} \circ D_{q_j}^{\text{phase}} \circ D_{\lambda_j}^{\text{depol}} \right)(\rho), \qquad (18)$$

where each $D$ represents a single-qubit channel with its corresponding Kraus operators, and $\circ$ denotes sequential composition of quantum channels. Physically, such dissipative dynamics arise from engineered couplings to local reservoirs or decay mechanisms, with parameters dependent on quantities like $T_1$ and $T_2$ relaxation times, thermal occupation, and environmental correlations. Each D-QPUF instance thus corresponds to a unique realization of these microscopic properties.

The evaluation of the QPUF proceeds through a quantum circuit composed of alternating unitary and dissipative layers. For an $n$-qubit system, the density matrix evolution at stage $k$ is given by

$$\rho_k = \Lambda^{(k)}\left(U_k \rho_{k-1} U_k^{\dagger}\right), \qquad (19)$$

where $U_k$ denotes the unitary operation at that stage and $\Lambda^{(k)}$ represents the local CPTP noise channel. Each noise map is defined by a tensor product of single-qubit Kraus operators, including amplitude damping (E), phase damping (F), and depolarization (G) channels. Sequential application of these noise processes models realistic decoherence phenomena observed in near-term quantum hardware. The cumulative effect across stages can be written as

$$\rho_{\text{final}} = \sum_{\alpha} \left(K_{\alpha_5}^{(5)} U_5 K_{\alpha_4}^{(4)} U_4 \cdots K_{\alpha_1}^{(1)} U_1\right) \rho_0 \left(K_{\alpha_1}^{(1)\dagger} U_1^{\dagger} \cdots K_{\alpha_5}^{(5)\dagger} U_5^{\dagger}\right), \qquad (20)$$

where $\alpha$ indexes the multi-stage Kraus operator sequence. The measurement outcome in the computational basis $|b\rangle\langle b|$ occurs with probability $Pr(b) = Tr(M_b \rho_{\text{final}})$. When classical readout noise is modelled by a transition matrix $R(b|b')$, the observed distribution becomes

$$p_{\text{meas}}(b) = \sum_{b'} R(b \mid b') \Pr(b'), \qquad (21)$$

allowing hardware-level imperfections to be incorporated into the response statistics.

The non-unitary nature of the D-QPUF follows directly from the structure of its CPTP map. A unitary operation maps pure states to pure states, while a dissipative process generally produces mixed outputs. For example, in the amplitude damping channel with Kraus operators based on equations 9 and 10, the action on the pure state $|1\rangle\langle 1|$ yields $E(|1\rangle\langle 1|) = (1-\gamma)|1\rangle\langle 1| + \gamma|0\rangle\langle 0|$, which is a mixed state. Since no unitary transformation can generate a mixed state from a pure input, the process is intrinsically non-unitary.

From a computational standpoint, the D-QPUF is demanding to simulate. If each qubit experiences $k_0$ Kraus operators, the total number of Kraus terms grows as $K_{\text{tot}} = k_0{}^{n}$. Applying this channel to a density matrix of dimension $d \times d$ requires $O(K_{\text{tot}} d^3)$ operations, leading to an asymptotic cost of $O\left((8k^0)^{n}\right)$. While this scaling is exponentially hard in $n$, significant reductions are achievable through vectorized simulations that apply local noise sequentially on the state vector. In such cases, each gate operation scales as $O(2^{n})$, and for a circuit with $G = O(n)$ gates, the total cost becomes $O(n2^{n})$. Memory requirements also scale exponentially: full storage of all challenge–response pairs for n-qubit basis challenges requires $\Theta(n2^{n})$ space.

Despite its complexity, the D-QPUF design provides a foundational realization of a physically grounded, non-unitary QPUF based on dissipative quantum channels. By coupling noise-induced irreversibility with measurement randomness, this architecture captures the essential properties of unforgeability and challenge–response diversity, while remaining physically implementable on noisy intermediate-scale quantum (NISQ) hardware.

### *4.2. Measurement-feedback Quantum Physical Unclonable Function*

A Quantum Physically Unclonable Function with measurement feedback (MF-QPUF) is a non-unitary quantum PUF architecture that integrates mid-circuit measurements and classically conditioned unitary gates. Unlike unitary designs, where the device's evaluation map is a reversible transformation, the MF-QPUF implements a full quantum instrument that combines measurement and feedback, producing an inherently non-unitary completely positive trace-preserving (CPTP) map.

Formally, the MF-QPUF is characterized by a triplet $(QGen, QEval, T)$. In the generation phase, $QGen(\lambda)$, given a security parameter $\lambda$, produces a device identifier that encodes a set of unitaries $\{U_1, \ldots, U_m\}$, projective measurement operators $\{M_b\}$, and conditional gates $\{V_b\}$ that are applied based on the outcomes of intermediate measurements. In the evaluation phase $QEval(id, \rho_{in})$, the input state $\rho_{in}$ first undergoes an initial unitary transformation $U_1$, producing $\rho_1 = U_1 \rho_{in} U_1^\dagger$. A subset of qubits is then measured using projectors $M_b$, yielding a classical outcome $b$ with probability $p_b = \mathrm{Tr}(M_b \rho_1)$. Based on this result, the conditional gate $V_b$ is applied to the remaining unmeasured qubits, and the process may iterate with additional measurement-feedback cycles. Since the intermediate classical outcomes are discarded, the overall channel describing the MF-QPUF output is a mixed quantum state,

$$\Lambda_{id}(\rho_{in}) = \sum_b V_b M_b U_1 \rho_{in} U_1^\dagger M_b V_b^\dagger, \quad (22)$$

which is a CPTP map that cannot be reduced to a unitary operator because of the non-reversible measurement process.

To describe the MF-QPUF circuit explicitly, consider an $n$-qubit challenge state $\rho_{in}$. After applying the initial unitary $U$, a subset of $m$ qubits is measured in the computational basis, producing a bit string $b \in \{0,1\}^m$. The corresponding projectors are $M_b = | b \rangle\langle b | \otimes I_{n-m}$. The intermediate state is $\rho' = U \rho_{in} U^\dagger$, the probability of outcome $b$ is $p_b = \mathrm{Tr}(M_b \rho')$, and the unnormalized post-measurement state is $\tilde{\rho}_b = M_b \rho' M_b$. Following conditional feedback, the branch state is $\rho_b = V_b \tilde{\rho}_b V_b^\dagger$. Since the outcome b is discarded, the complete transformation is based on equation 22, which defines a CPTP, but generally non-unitary channel.

As a concrete example, consider an MF-QPUF circuit comprising four qubits labelled $i \in \{0,1,2,3\}$, each subject to distinct decoherence ($\gamma_i$), phase damping ($p_i^{(\phi)}$), and depolarization ($p_i^{(d)}$) parameters. Noise is modelled via Kraus operators $K_{a,p,d}^{(i)} = D_d^{(i)} P_p^{(i)} A_a^{(i)}$ with indices $a \in \{0,1\}, p \in \{0,1\}$ and $d \in \{0,1,2,3\}$, and the total operator for the system is $K_\alpha = \otimes_{i=0}^{3} K_{a_i,p_i,d_i}^{(i)}$, where $\alpha = (a_0, p_0, d_0; \dots; a_3, p_3, d_3)$.

The circuit proceeds as follows. The four-qubit input state $\rho_0$ first undergoes the initial sequence $U_{\mathrm{init}} = H_0 \otimes (H_1 X_1) \otimes H_2 \otimes H_3$, resulting in $\rho_A = U_{\mathrm{init}} \rho_0 U_{\mathrm{init}}^\dagger$, followed by noise $N^{(A)}$. The entangling stage applies $U_{\mathrm{ent1}} = \mathrm{CNOT}_{1\to 0}$ to produce $\rho_B = U_{\mathrm{ent1}} \rho_A^{\mathrm{noisy}} U_{\mathrm{ent1}}^\dagger$, then $U_{\mathrm{ent2}} = \mathrm{CNOT}_{2\to 1}$ to yield $\rho_C^{\mathrm{noisy}}$. A mid-circuit measurement is then performed on qubit 2 using projectors $\Pi_0^{(2)} = | 0 \rangle\langle 0 |$ and $\Pi_1^{(2)} = | 1 \rangle\langle 1 |$, extended to four qubits as $M_0 = I_0 \otimes I_1 \otimes \Pi_0^{(2)} \otimes I_3$ and $M_1 = I_0 \otimes I_1 \otimes \Pi_1^{(2)} \otimes I_3$. The unnormalized branch state is $\sigma_m = M_m \rho_C^{\mathrm{noisy}} M_m$ with probability $p_m = \mathrm{Tr}(\sigma_m)$, yielding the normalized state $\rho_m = \sigma_m / p_m$. Conditional gate $V_m$ is applied to each branch, resulting in the map

$$\mathcal{E}_m(\rho_0) = \sum_\alpha L_{(m,\alpha)} \rho_0 L_{(m,\alpha)}^\dagger, L_{(m,\alpha)} = V_m M_m K_\alpha^{(C)} U_{\mathrm{ent2}} U_{\mathrm{ent1}} U_{\mathrm{init}}. \quad (23)$$

After additional post-processing through $U_{\mathrm{post}}^{(m)}$ and noise $N^{(\mathrm{post})}$, the final output is expressed as

$$\rho_{\mathrm{out}} = \sum_{m\in\{0,1\}} \sum_\beta \sum_\alpha G_{(m,\beta,\alpha)} \rho_0 G_{(m,\beta,\alpha)}^\dagger, \quad (24)$$

where the overall Kraus operators are

$$G_{(m,\beta,\alpha)} = K_\beta^{(\text{post})} U_{\text{post}}^{(m)} V_m M_m K_\alpha^{(C)} U_{\text{ent2}} U_{\text{ent1}} U_{\text{init}}. \quad (25)$$

This generalizes naturally to multiple mid-circuit measurements: for a subset $S \subseteq \{0,1,2,3\}$ of measured qubits, the outcome $m_S \in \{0,1\}^{|S|}$ defines projectors $M_{(m_S)} = \otimes_{i\in S}\Pi_{m_i}^{(i)} \otimes \otimes_{j\notin S} I^{(j)}$, conditional gates $U_{(m_S)}$, and branch maps $\mathcal{I}_{(m_S)}(\rho) = U_{(m_S)} M_{(m_S)} \rho M_{(m_S)} U_{(m_S)}^\dagger$. Including noise yields complete Kraus sets $G_{(m_S,a)}$ and the global transformation

$$\rho_{\text{out}} = \sum_{m_S} \sum_a G_{(m_S,a)} \rho_0 G_{(m_S,a)}^\dagger \quad (26)$$

The non-unitary character of the MF-QPUF follows directly from this formulation. The map

$$\mathcal{E}(\rho) = \sum_m U_m M_m \rho M_m^\dagger U_m^\dagger \quad (27)$$

represents a probabilistic mixture of quantum trajectories. For instance, measuring a single qubit with projectors $M_0 =| 0\rangle\langle 0 |, M_1 =| 1\rangle\langle 1 |$ and applying $X$ or $I$ depending on the results yields $\mathcal{E}(\rho) = U_0 M_0 \rho M_0^\dagger U_0^\dagger + U_1 M_1 \rho M_1^\dagger U_1^\dagger$. If $\rho =| +\rangle\langle +|$ with $| +\rangle = (| 0\rangle + | 1\rangle)/\sqrt{2}$, the result is a mixed state that cannot be generated by any unitary transformation. The measurement introduces irreversible decoherence between branches, proving the intrinsic non-unitarity of the MF-QPUF mapping.

In terms of computational complexity, the MF-QPUF exhibits two major sources of exponential growth: the dimensionality of the Hilbert space and the branching structure induced by feedback. If each measurement produces $b$ possible outcomes and there are $f$ feedback rounds, the number of trajectories scales as $b^f$ Denoting the cost of simulating a single trajectory by $\text{Cost}_{\text{traj}}^{\text{MF}}(n, m)$, the cost of estimating the full output distribution with $S$ monte carlo samples is $\text{Cost}_{\text{MC}}^{\text{MF}}(n, m, f, \varepsilon) = O(b^f S(\varepsilon)\text{Cost}_{\text{traj}}^{\text{MF}}(n, m))$. For binary measurements ($b = 2$) and $S(\varepsilon) = \Theta(1/\varepsilon^2)$, this becomes

$$\text{Cost}_{\text{MC}}^{\text{MF}}(n, m, f, \varepsilon) = \Theta\left(\frac{2^f}{\varepsilon^2}\text{Cost}_{\text{traj}}^{\text{MF}}(n,m)\right).$$

If matrix-based simulation is employed, each measurement branch must be computed separately, increasing the cost by approximately $2^f$. For trajectories scaling as $O(m \cdot \text{poly}(n) \cdot 2^n)$, the overall cost grows as $O((2^f m\, \text{poly}(n) 2^n)/\varepsilon^2)$. Tomographic characterization of the MF-QPUF similarly incurs a higher sample complexity due to the branching structure, potentially reaching $O((2^f d^4)/\varepsilon^2)$.

Altogether, the MF-QPUF introduces an intrinsically adaptive and non-unitary layer to the QPUF paradigm, with richer statistical behaviour and greater expressive power but also substantially higher simulation and characterization costs.

*4.3. Linblad Quantum Physical Unclonable Function*

The Lindblad-based Quantum Physically Unclonable Function (L-QPUF) is defined as a device whose internal response is governed by engineered Lindblad dynamics, combining dissipative evolution and unitary entanglement in a controlled non-unitary channel. The secret identifier $id$ of the device encodes a hidden sequence of entangled unitary gates $U_1, \ldots, U_m$, linblad generators $L_1, \ldots, L_m$ together with their corresponding jump operators, decay rates $\{\gamma_{(j,k)}\}$, and interaction times $t_1, \ldots, t_m$. The overall evaluation channel associated with this identifier is the layered composition

$$\Lambda_{id} = (e^{L_m t_m} \circ U_m) \circ \cdots \circ (e^{L_1 t_1} \circ U_1), \qquad (28)$$

where each $U_k(\rho) = U_k \rho U_k^\dagger$ and, for an input challenge state $\rho_{\text{in}}$, the device outputs $\rho_{\text{out}} = \Lambda_{id}(\rho_{\text{in}})$. This mapping is completely positive and trace-preserving (CPTP) yet non-unitary, since each $e^{L_k t_k}$ represents open quantum evolution under a Lindblad generator. The Lindblad form represents a Markovian process—derived under the Born–Markov approximation—where the system has no effective memory of the environment, producing a dynamical semigroup that satisfies CPTP and time-divisibility properties, such that $e^{Lt} = (e^{L\Delta})^r$.

For implementation, each dissipative block $e^{L_k t_k}$ is decomposed into smaller steps $\Delta_k = t_k / r_k$ and approximated using either the Lie-Trotter or Strang expansion, yielding

$$e^{L_k t_k} \approx \Big( \prod_{\alpha \in \text{block } k} e^{L_{(k,\alpha)} \Delta_k} \Big)^{r_k}. \qquad (29)$$

If $L = \sum_l L_l$, the first-order Lie–Trotter approximation $(\prod_l e^{L_l \Delta})^r \to e^{Lt}$ achieves error

$$O(t^2 / r \cdot \max_{l<k} \| [L_l, L_k] \|_\diamond), \qquad (30)$$

while the second-order Strang expansion $(e^{L_1 \Delta/2} \cdots e^{L_s \Delta} \cdots e^{L_1 \Delta/2})^r \to e^{Lt}$ yields a higher-order error

$$O(t^3 / r^2 \cdot \max \| [L, [L', L'']] \|_\diamond). \qquad (31)$$

Thus, using Strang blocks with sufficient $r_k$ ensures both accuracy and preservation of CPTP structure at every step. Importantly, this convergence and the physical positivity of intermediate states are guaranteed by the Markovian semigroup property of the generator.

To connect the Lindblad and Kraus pictures for a small-time step $\Delta t$, consider the generator

$$L(\rho) = -i[H, \rho] + \sum_j \gamma_j \left( L_j \rho L_j^\dagger - \frac{1}{2} \{ L_j^\dagger L_j, \rho \} \right), \qquad (32)$$

whose first-order expansion gives

$$e^{L\Delta t}(\rho) = K_0 \rho K_0^\dagger + \sum_j K_j \rho K_j^\dagger + O(\Delta t^2), \qquad (33)$$

with

$$K_j = \sqrt{\gamma_j \Delta t}\, L_j \qquad (34)$$

and

$$K_0 = I - iH\Delta t - \frac{1}{2} \sum_j \gamma_j L_j^\dagger L_j \Delta t \qquad (35)$$

This set of Kraus operators approximates a CPTP map to $O(\Delta t^2)$ and becomes exactly CPTP under higher-order corrections or full exponentiation. However, this argument holds precisely when $L$ is time-independent and Markovian; time-dependent or non-Markovian generators require either explicit environmental modeling or positivity-preserving alternatives. The detailed treatment of the Trotter–Suzuki approximation is presented in Appendix A.

An example of circuit realization can be illustrated with a four-qubit L-QPUF. The initial state $\rho_0$ undergoes the gate sequence

$$U_{\text{init}} = H^{(0)} \otimes \left(H^{(1)}X^{(1)}\right) \otimes H^{(2)} \otimes H^{(3)}, \tag{36}$$

producing $\rho_1 = U_{\text{init}}\rho_0 U_{\text{init}}^{\dagger}$. The system then experiences a linblad window $e^{L^{(1)}\tau_1}$, followed by an entangling gate $U_{\text{ent1}} = \text{CX}_{1\to 0}$, another dissipative window $e^{L^{(2)}\tau_2}$, a local rotation layer $U_{\text{local}}$, and successive dissipative and entangling stages $e^{L^{(3)}\tau_3}$, $U_{\text{ent2}}$, $e^{L^{(4)}\tau_4}$, and finally a concluding rotation $U_{\text{final}}$ with terminal evolution $e^{L^{(5)}\tau_5}$. Projective measurements in the computational basis with projectors $\Pi_0^{(i)}, \Pi_1^{(i)}$ yield outcomes $m = (m_0, m_1, m_2, m_3)$ and corresponding probabilities $p_m = \text{Tr}(M_m \rho_5' M_m^{\dagger})$, where $M_m = \otimes_i \Pi_{m_i}^{(i)}$.

To express the path-operator formulation, each Lindblad window can be decomposed into elementary operators $T_{(\alpha;t)}^{(l)}$, allowing the complete path operator for branch $m$ to be written as

$$G_{(m;\{\alpha^{(l)}\})} = M_m \left( \prod_{\text{final } l} [N^{(l)}(\tau_l) \prod_j L_{\left(\alpha_j^{(l)}\right)}^{(l)}] \right) U_{\text{final}} \cdots \left( \prod_{\text{first } l} [N^{(l)}(\tau_l) \prod_j L_{\left(\alpha_j^{(l)}\right)}^{(l)}] \right) U_{\text{init}}. \tag{37}$$

Integrating over all possible jump sequences and times yields the final state

$$\rho_{\text{out}} = \sum_m \sum_{\{\alpha^{(l)}\}} \int G_{(m;\{\alpha^{(l)}\})} \rho_0 G_{(m;\{\alpha^{(l)}\})}^{\dagger}. \tag{38}$$

Each local decay process is embedded into the global space via tensoring with identities. For instance, amplitude damping on qubit $i$ is

$$L_{\text{AD}}^{(i)} = \sqrt{\gamma_{\text{AD}}^{(i)}} (I^{\otimes i} \otimes |\, 0\rangle\langle 1 \,| \otimes I^{\otimes(n-i-1)}); \tag{39}$$

dephasing is

$$L_{\phi}^{(i)} = \sqrt{\gamma_{\phi}^{(i)}} (I^{\otimes i} \otimes Z \otimes I^{\otimes(n-i-1)}); \tag{40}$$

and depolarization channels use $L_X^{(i)}, L_Y^{(i)}, L_Z^{(i)}$ with corresponding rates $\kappa^{(i)}$. Collective and pairwise dissipation processes are expressed respectively as

$$L_{\text{coll}}^{(A)} = \sqrt{\Gamma_A} \sum_i c_i \left(I^{\otimes i} \otimes A \otimes I^{\otimes(n-i-1)}\right), \tag{41}$$

and

$$L_{\text{pair}}^{(B)} = \sqrt{\Gamma_B}\sum_{i,j} c_{ij}(I^{\otimes i} \otimes B \otimes I^{\otimes(j-i-1)} \otimes B \otimes I^{\otimes(n-j-1)}), \tag{42}$$

for $A, B \in \{X, Y, Z\}$.

In the discretized picture, each small-step effective Hamiltonian is

$$H_{\text{eff}} = H - \frac{i}{2}\sum_{\alpha} L_\alpha^\dagger L_\alpha, \tag{43}$$

with Kraus operators $K_0 = I - iH_{\text{eff}}\Delta t$ and $K_\alpha = \sqrt{\Delta t} L_\alpha$. Thus, every Trotter step consists of one no-jump operator and a set of jump operators defining a physically valid CPTP map.

The overall evaluation map can be written as equation 38.

where each $G_{(m_S,a)}$ combines the effects of measurements, feedback, and dissipative noise. The randomization of Lindblad rates $\{\gamma_j\}$ across queries enhances unforgeability, ensuring that no adversary can aggregate or infer query correlations to predict unseen outputs.

The L-QPUF therefore realizes a challenge–response mapping through a layered structure of entangled unitaries and engineered Lindblad evolutions. Its path-integral representation $G_{(m;\{\alpha\})}$ encapsulates all trajectories, the Kraus expansions control numerical accuracy, and the Trotter–Strang decomposition ensures stable CPTP evolution. From a security standpoint, the non-unitary nature of this model, along with its stochastic rate randomization, renders efficient adversarial learning of the channel $\Lambda_{id}$ computationally infeasible.

The non-unitarity of the L-QPUF follows directly from the structure of the Lindblad equation (equation 32). If only the commutator term were present, the evolution would be $\rho(t) = U(t)\rho(0)U(t)^\dagger$, i.e., unitary. However, the additional dissipative terms cause entropy growth and information leakage into the environment, forming an irreversible CPTP semigroup $e^{tL}$ rather than a reversible unitary operator. As an illustrative example, the amplitude-damping generator $L = \sqrt{\gamma} \mid 0\rangle\langle 1 \mid$ acting on the pure state $\mid 1\rangle\langle 1 \mid$ produces a mixed output, which cannot result from any unitary process. More generally, by the fundamental theorem of quantum dynamical semigroups, any CPTP semigroup evolution is necessarily non-unitary, establishing the theoretical foundation for the irreversibility of the L-QPUF design.

Regarding computational cost, let each Lindblad layer contain $K_{\text{step}}$ Kraus operators. The naïve cost of applying $m$ layers on a $d$-dimensional system scales as $O(m\, K_{\text{step}}\, d^3)$. For a localized model with $k_0$ operators per qubit and $n$ qubits, this yields $O(m(8k_0)^n)$. In gate-based simulation, decomposition of each step into elementary rotations or auxiliary qubits reduces the per-step cost to $\tilde{O}(d)$ - $\tilde{O}(d^2)$, depending on whether the full density matrix or stochastic trajectories are simulated. In the trajectory-based scheme, the cost for $S$ sampled trajectories is $O(Sm\,\text{poly}(n)\, 2^n)$. Process tomography of the L-QPUF requires $N_{\text{proc}}^{\text{LD}}(\varepsilon) = \Theta(d^4/\varepsilon^2) = \Theta(16^n/\varepsilon^2)$ samples in general; with low-rank assumptions ($r \ll d^2$), compressed tomography reduces this to $\tilde{O}(rd^2/\varepsilon^2)$, though such knowledge is unrealistic for an adversary. Storing all challenge–response pairs demands $\Theta(n2^n)$ space.

In essence, the L-QPUF architecture merges entangled unitary operations with dissipative Lindblad dynamics to construct a cryptographically secure, physically unclonable map. Its non-unitary semigroup nature ensures intrinsic irreversibility, while controlled randomization of parameters maintains unpredictability and resistance to quantum or classical attacks.

### *4.4. Comparison of the three designs*

The three non-unitary QPUF architectures—D-QPUF, MF-QPUF, and L-QPUF—represent distinct operational realizations of open-system dynamics for implementing physically unclonable functions on quantum hardware.

Although each architecture follows the same abstract structure of a challenge–response evaluation map $\rho_{\text{out}} = \mathcal{E}(\rho_{\text{in}})$, their internal mechanisms, physical assumptions, and complexity–security trade-offs differ fundamentally. A comparative analysis clarifies the design space and highlights the complementary advantages of each approach.

The dissipative QPUF (D-QPUF) constitutes the simplest instantiation, in which the non-unitarity arises from engineered amplitude damping or dephasing channels applied to a small set of qubits. Because its dynamics can be represented by low-rank Kraus operators, D-QPUF achieves high hardware efficiency and low sampling cost. The physical entropy production due to energy relaxation ensures randomness and partial irreversibility, while the relatively limited non-unitarity bounds the challenge-response diversity. Consequently, D-QPUF offers strong reproducibility and stability but a lower entropy density compared with the more complex designs. It is therefore suited to near-term devices and low-overhead authentication scenarios.

In contrast, the measurement-feedback QPUF (MF-QPUF) introduces an explicit measurement-conditioned control loop. Mid-circuit measurements extract partial classical information that dynamically influences subsequent unitaries, transforming the evolution into a measurement-driven stochastic process. The hybrid quantum–classical feedback chain increases effective non-unitarity and enables adaptive amplification of microscopic fluctuations into macroscopically distinct outcomes. While this design achieves a much larger challenge space and higher min-entropy per response, it incurs additional latency from real-time feedback and higher classical memory requirements for storing outcome-conditioned control histories. The dependence on low-latency measurement and feedforward fidelity makes MF-QPUF best suited to hardware platforms supporting fast mid-circuit readout and digital control, such as superconducting or trapped-ion processors.

The Lindblad-based QPUF (L-QPUF) generalizes the concept by embedding the QPUF evaluation within a continuous-time open-system evolution governed by a generator $L$ of Lindblad form. This model captures both dissipative and dephasing mechanisms and allows controllable randomization of decay rates $\{\gamma_j\}$ as hidden device parameters. As a result, L-QPUF exhibits the strongest theoretical unforgeability: reproducing its response distribution requires complete knowledge of the environmental coupling operators and decoherence rates, information inherently inaccessible to an adversary. However, these advantages come with significantly higher simulation and realization costs, since evaluating $e^{Lt}$ requires exponential resources in the number of jump operators and time slices. L-QPUF thus represents the most physically faithful but computationally intensive member of the QPUF family. The computational resources required for each proposed design, including total gate count, simulation cost, and tomography overhead, are presented in Table 2 The results highlight the relative efficiency of the D-QPUF, MF-QPUF, and L-QPUF under increasing qubit numbers.

***Table 2.*** *Results of complexity analysis of gate count, simulation cost, and tomography cost for the introduced designs*

| Design/Cost | #Gates | Tomography | Simulation |
|---|---|---|---|
| **D-QPUF** | $O(n)$ | $\mathrm{O}(16^n/\varepsilon^2)$ | $O(n2^n)$ |
| **MF-QPUF** | $O(m \cdot poly(n))$ | $O((2^f \cdot 16^n)/\varepsilon^2)$ | $O(2^f \cdot m \cdot poly(n) \cdot 2^n/\varepsilon^2)$ |
| **L-QPUF** | $O(m \cdot poly(n))$ | $O(16^n/\varepsilon^2)$ | $O(m \cdot poly(n) \cdot 2^n)$ |

From a comparative standpoint, the three architectures trace an ascending hierarchy in both physical completeness and computational cost. D-QPUF captures basic dissipative randomness with minimal hardware overhead; MF-QPUF extends this with measurement-feedback-induced stochasticity, achieving higher entropy and adaptive control; L-QPUF provides the full open-quantum-system formulation with maximal security at the expense of efficiency. Quantitatively, the sample complexity scales approximately as $\mathrm{O}(2^{\mathrm{n}})$, $\mathrm{O}(2^{\mathrm{n}} \cdot \mathrm{m})$, and $\mathrm{O}(2^{\mathrm{n}} \cdot$ $m^r$) for the three designs, respectively, where n denotes qubit number, mmm the number of feedback rounds, and r the number of time slices in the Lindblad decomposition. Correspondingly, response entropy and unforgeability increase along the same sequence.

In practice, these designs need not be mutually exclusive. Hybrid schemes—such as dissipative channels interleaved with measurement feedback under Lindbladian supervision—may further balance the trade-off between reproducibility, entropy, and computational overhead. The comparative study thus delineates a scalable continuum of non-unitary QPUF realizations, from experimentally lightweight dissipative prototypes to theoretically rigorous open-system constructions, establishing a foundation for selecting or combining architectures according to the available quantum hardware and desired security level.

## 5. Simulation results

This section presents the simulation results of the three proposed non-unitary Quantum Physical Unclonable Function (QPUF) designs—D-QPUF, MF-QPUF, and L-QPUF—and evaluates their behavior under both ideal and noisy conditions. The primary performance metrics include uniformity, uniqueness, and reliability, which collectively measure the statistical balance of output bits, the distinguishability of device instances, and the consistency of responses under repeated challenges. Each design was tested on circuits comprising two to eight qubits, implemented and executed using the Qiskit framework [39]. The experiments were first conducted on the Qiskit Aer simulator to examine ideal, noise-free conditions and subsequently validated on three IBM quantum processors—Athens, Santiago, and Melbourne—to assess real-world feasibility and robustness against hardware-induced noise. Additionally, the trends of the characteristic hardware parameters $T_1$, $T_2$, and readout error rates were examined as functions of the number of circuit executions to analyze the long-term stability of QPUF responses.

### *5.1. Simulation environment*

To evaluate the proposed QPUF architectures, both local quantum circuit simulators and hardware-based backends were utilized. Local simulators such as Qiskit Aer execute quantum circuits on classical hardware, allowing experiments in an idealized, noise-free environment. This enables direct analysis of theoretical properties, entropy measures, and circuit stability without the stochastic influence of decoherence or readout errors. By contrast, backend simulations model or directly use real quantum hardware, where imperfections such as gate infidelity, limited coherence times, and restricted connectivity introduce genuine noise into the computation. The complementary use of both approaches ensures that the proposed designs are tested under ideal theoretical conditions and realistic noisy hardware scenarios representative of current NISQ technology.

The Qiskit Aer simulator was chosen as the primary local simulator due to its support for both statevector and density-matrix representations, as well as its configurable noise models. The density-matrix formalism is particularly important for accurately simulating non-unitary processes such as dissipation, dephasing, and mid-circuit measurement effects. Aer thus provides a high-fidelity platform for validating the correctness of QPUF implementations before hardware deployment.

For hardware validation, three superconducting quantum processors were selected: IBM Athens, IBM Santiago, and IBM Melbourne. Although these devices have been decommissioned, Qiskit continues to provide *simulated backends* based on their final calibration data, including measured parameters such as $T_1$, $T_2$, multi-qubit gate error rates, and readout errors. These backends effectively reproduce the behaviour of the original hardware, enabling realistic benchmarking of QPUF designs without the need for live hardware access.

The Athens processor features a linear chain topology of five qubits, suitable for small, shallow circuits but requiring multiple SWAP operations for larger ones, thereby increasing circuit depth and error probability. The Santiago processor, also a five-qubit device, employs a star topology with one central qubit connected to all others, enabling efficient two-qubit operations but with asymmetric noise distribution. The Melbourne processor extends to fifteen qubits arranged in a ladder-like topology, designed to explore deeper circuits and correlated errors, providing insight into scalability effects. The physical qubit connectivity maps of the IBM Athens, Santiago, and Melbourne quantum devices used in this work are illustrated in Figures 1–3, respectively. These topologies determine the available two-qubit couplings and, consequently, the routing efficiency of our QPUF circuits.

***Figure 1.*** *Athens Quantum computer qubits topology*

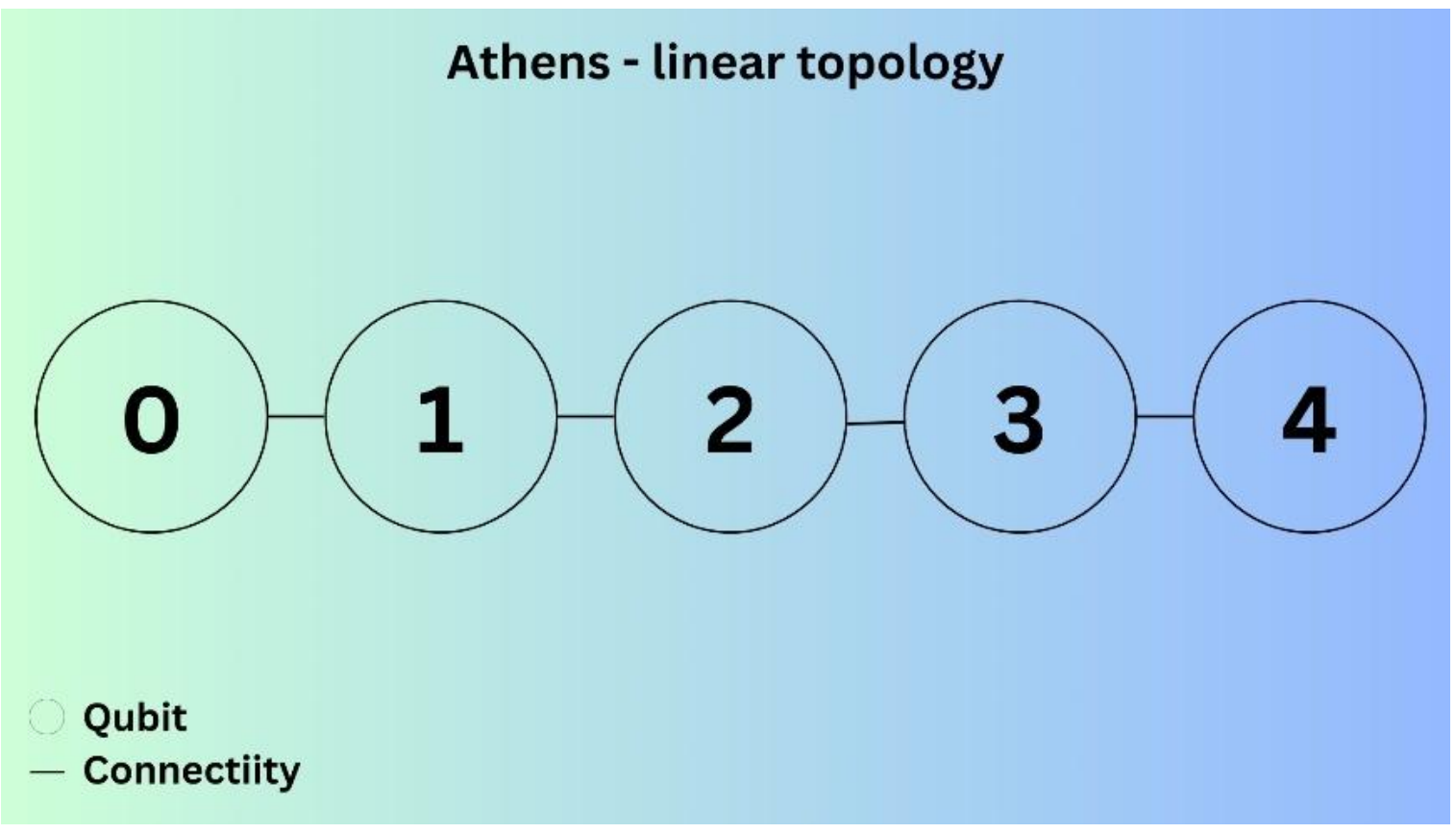

***Figure 2.*** *Santiago Quantum computer qubits topology*

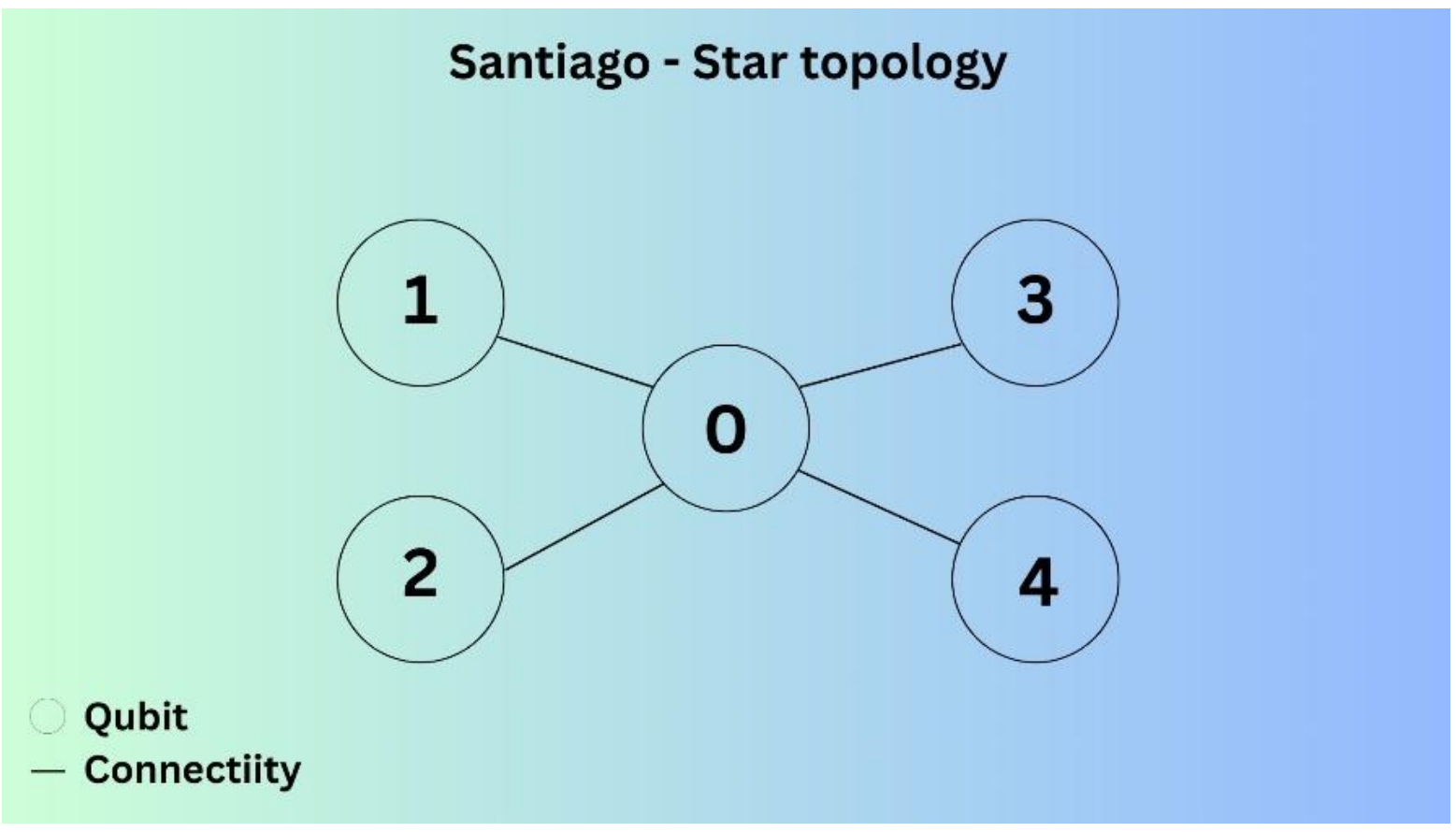

The choice of these three processors was motivated by three main considerations. First, they represent historically well-studied platforms that have been extensively used in QPUF and noise-resilient algorithm research, facilitating comparison with prior work. Second, their topological diversity and differing qubit counts allow a systematic evaluation of how connectivity and system scale influence QPUF behaviour. Third, their relatively higher noise levels and parameter fluctuations provide a conservative, “worst-case” testbed, useful for assessing the robustness of the proposed designs. In combination, these simulators and backends provide a comprehensive testing environment that bridges theoretical analysis and practical feasibility.

***Figure 3.*** *Melbourne Quantum computer qubits topology*

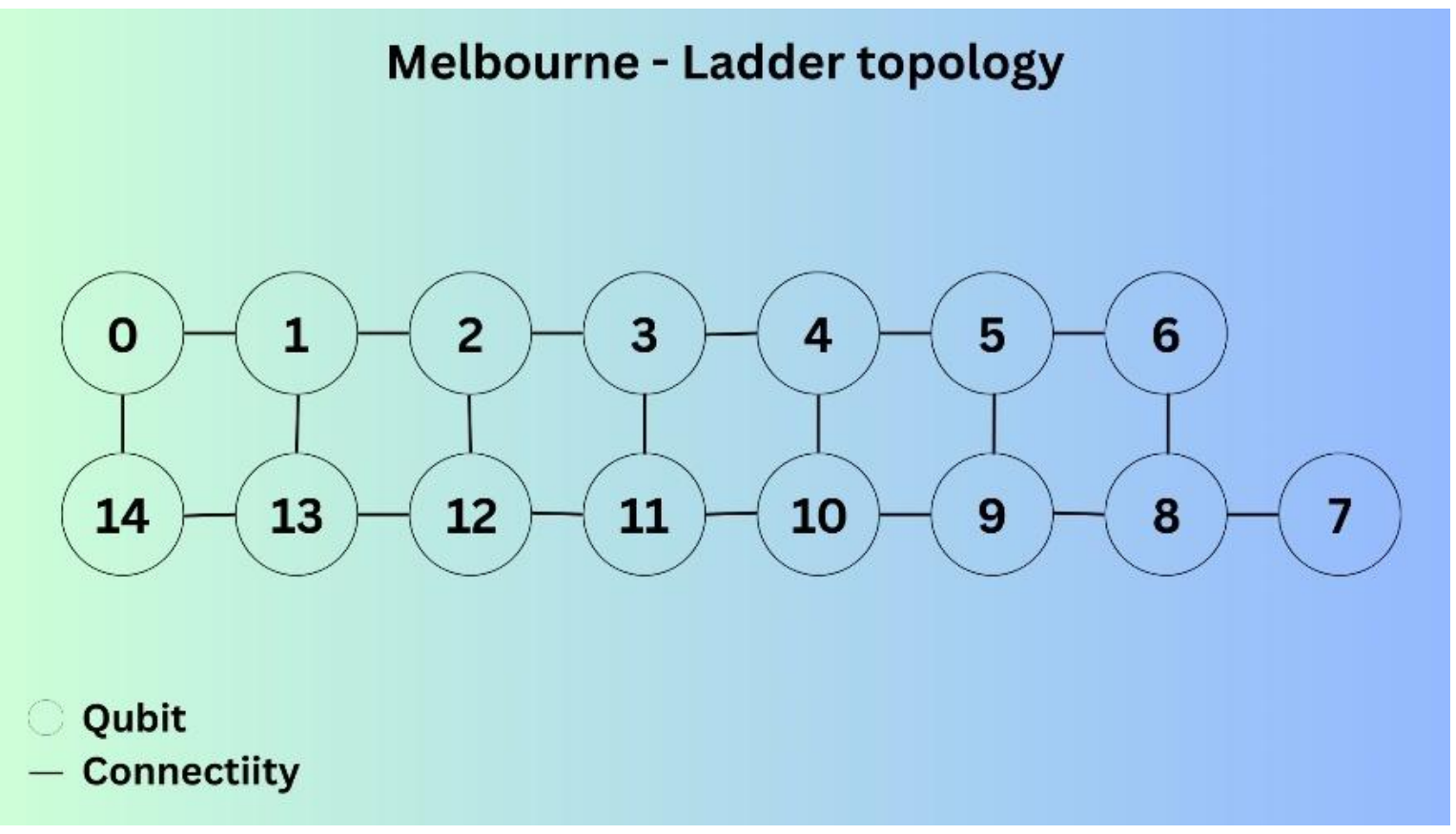

*5.2. Simulation parameters*

The evaluation of non-unitary QPUF designs involves several key parameters that control circuit complexity, randomness, and response diversity. Among them, the parameters $m$ and $f$ play particularly critical roles in the L-QPUF and MF-QPUF models, respectively, as they determine the depth of non-unitary evolution and the number of feedback cycles applied within the circuit.

In the L-QPUF, mmm denotes the number of Lindblad map steps applied sequentially to the quantum state. Each step corresponds to a non-unitary dissipative transformation that progressively enhances randomness and irreversibility. When $m = 1$, the circuit behaves similarly to a simple noisy channel, offering limited unpredictability and reduced security. For $m \geq 2$, the cumulative effect of successive Lindblad steps significantly amplifies the diversity of challenge–response pairs (CRPs). However, larger values of $m$ increase $m = 2$ was selected as a balance between expressivity and efficiency, with additional tests up to $m = 3$ conducted to investigate deeper non-unitary evolution.

In the MF-QPUF, $f$ represents the number of measurement-feedback cycles, each involving a mid-circuit measurement followed by conditional application of unitary gates based on the classical outcome. These feedback cycles introduce branching evolution paths, exponentially increasing the response diversity and resistance to cloning or emulation. When $f = 0$, the circuit reduces to the D-QPUF model, losing its adaptive stochasticity. While higher $f$ enhances security, it also raises the number of classical–quantum interactions and simulation complexity. In this work, $f = 1$ was adopted as the default, with $f = 2$ explored for extended analysis. It was observed that additional feedback substantially increased simulation time due to the exponential growth of measurement-conditioned trajectories.

Beyond these structural parameters, the simulations employed consistent experimental settings for all designs to enable fair comparison. The number of qubits ranged from two to eight, depending on the design and hardware constraints. For each design, 50 independent QPUF instances were generated to emulate different physical realizations, and 100 random challenges were applied to each instance. Each circuit was executed 10,000 times to obtain statistically reliable output distributions. The employed noise models included decoherence, dephasing, and depolarization, with noise coefficients drawn uniformly from the interval $[0.001, 0.05]$, covering the range from nearly ideal to moderately noisy hardware conditions.

QPUF responses were extracted from measurement outcomes using a majority voting approach, ensuring deterministic response generation despite probabilistic measurement noise. This method provides consistent binary outputs across repeated evaluations of the same challenge, a property essential for both reliability analysis and real-world authentication. Under these settings, the three QPUF architectures were systematically evaluated for uniformity, uniqueness, and reliability, providing quantitative insights into their comparative behaviour and practical feasibility.

### *5.3. Analysis of Run-time Noise in Quantum Computers*

This section analyses the noise characteristics and coherence parameters obtained from the three tested IBM Quantum backends during the simulation process. Table 3 reports the number of executed samples, average values, minimum and maximum means, maximum range among $T_1$ and $T_2$, standard deviation, and readout error for each backend. Device-level noise characteristics are visualized in Figures 4–12, which show the frequency distributions of $T_1$, $T_2$, and readout error across all physical qubits for each quantum computer. These results reveal substantial variation in coherence and measurement fidelity, motivating the adoption of non-unitary models in our designs. The relaxation and dephasing times ($T_1$ and $T_2$) are measured in microseconds, while the readout error is expressed as a percentage. Generally, higher $T_1$ and $T_2$ values combined with lower readout errors indicate a more noise-resilient and reliable quantum processor.

***Table 3.*** *Results of $T_1$, $T_2$, $T_\theta$, and readout error for three quantum computers during execution. Times are reported in microseconds, and readout error is reported in percentage.*

| **Backend** | **Athens** | | | | **Santiago** | | | | **Melbourne** | | | |
|---|---|---|---|---|---|---|---|---|---|---|---|---|
| **Metric** | $T_1$ | $T_2$ | $T_\theta$ | **Readout error** | $T_1$ | $T_2$ | $T_\theta$ | **Readout error** | $T_1$ | $T_2$ | $T_\theta$ | **Readout error** |
| **Mean** | 75.78 | 90.46 | 224.39 | 2.45 | 133.14 | 123.33 | 229.73 | 1.95 | 55.02 | 59.87 | 131.32 | 7.02 |
| **Min** | 57.47 | 50.98 | 91.61 | 1.27 | 83.59 | 55.7 | 83.53 | 0.98 | 47.78 | 44.39 | 85.7 | 4.1 |
| **Max** | 103.87 | 125.23 | 315.3 | 24.06 | 163.19 | 165.09 | 334.07 | 5.59 | 60.45 | 70.79 | 211.7 | 10.16 |
| **Range** | 46.4 | 74.25 | 371.45 | 22.79 | 79.61 | 109.39 | 349.53 | 4.61 | 12.68 | 26.4 | 126 | 6.07 |
| **Std Dev** | 8.03 | 12.68 | 60.25 | 2.93 | 13.63 | 17.01 | 45.24 | 0.77 | 2.04 | 4.73 | 55 | 1.09 |
| **Samples** | 66747 | | | | 74576 | | | | 47112 | | | |

The results show that the Melbourne backend exhibits an average readout error of 7.02%, significantly higher than those of Athens and Santiago, implying that its measurement outcomes are less reliable and more affected by noise. In contrast, Santiago demonstrates the lowest average readout error of 1.95%, followed by Athens with 2.45%, indicating that Santiago is the most accurate in terms of measurement fidelity.

Regarding energy relaxation ($T_1$), Santiago again achieves the best performance with an average of 133.14 $\mu s$, followed by Athens with 75.78 $\mu s$, and Melbourne with only 55.02 $\mu s$. This suggests that qubits on Santiago maintain their excited states longer before relaxing, resulting in greater operational stability.

Similarly, for phase coherence ($T_2$), Santiago leads with an average of 123.33 $\mu s$, while Athens records 90.46 $\mu s$ and Melbourne 59.87 $\mu s$. Thus, Santiago demonstrates superior preservation of quantum coherence, making it the most favorable backend for quantum operations requiring high fidelity.

***Figure 4.*** *Frequency distribution of $T_1$ values per physical qubit on Athens*

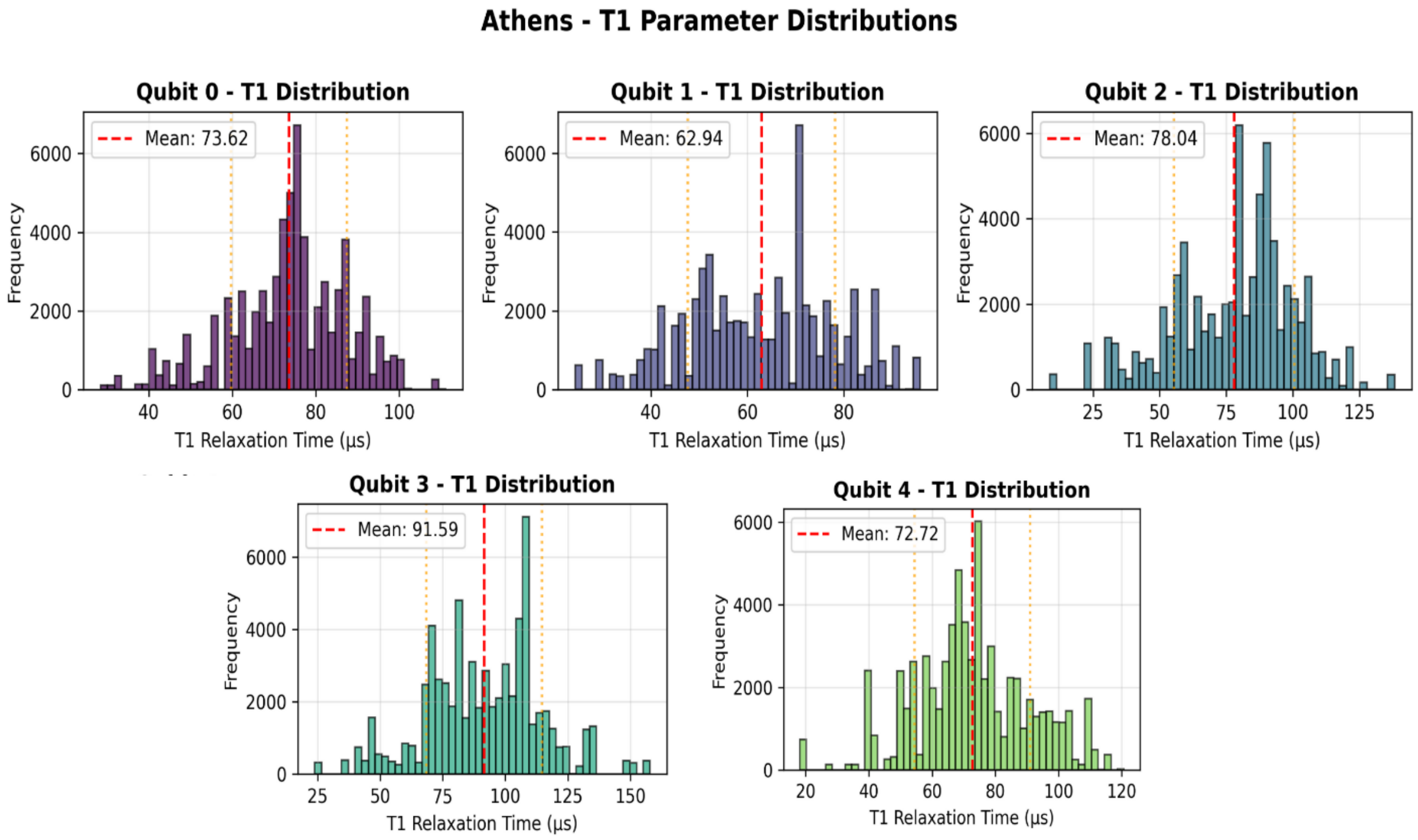

***Figure 5.*** *Frequency distribution of $T_2$ values per physical qubit on Athens*

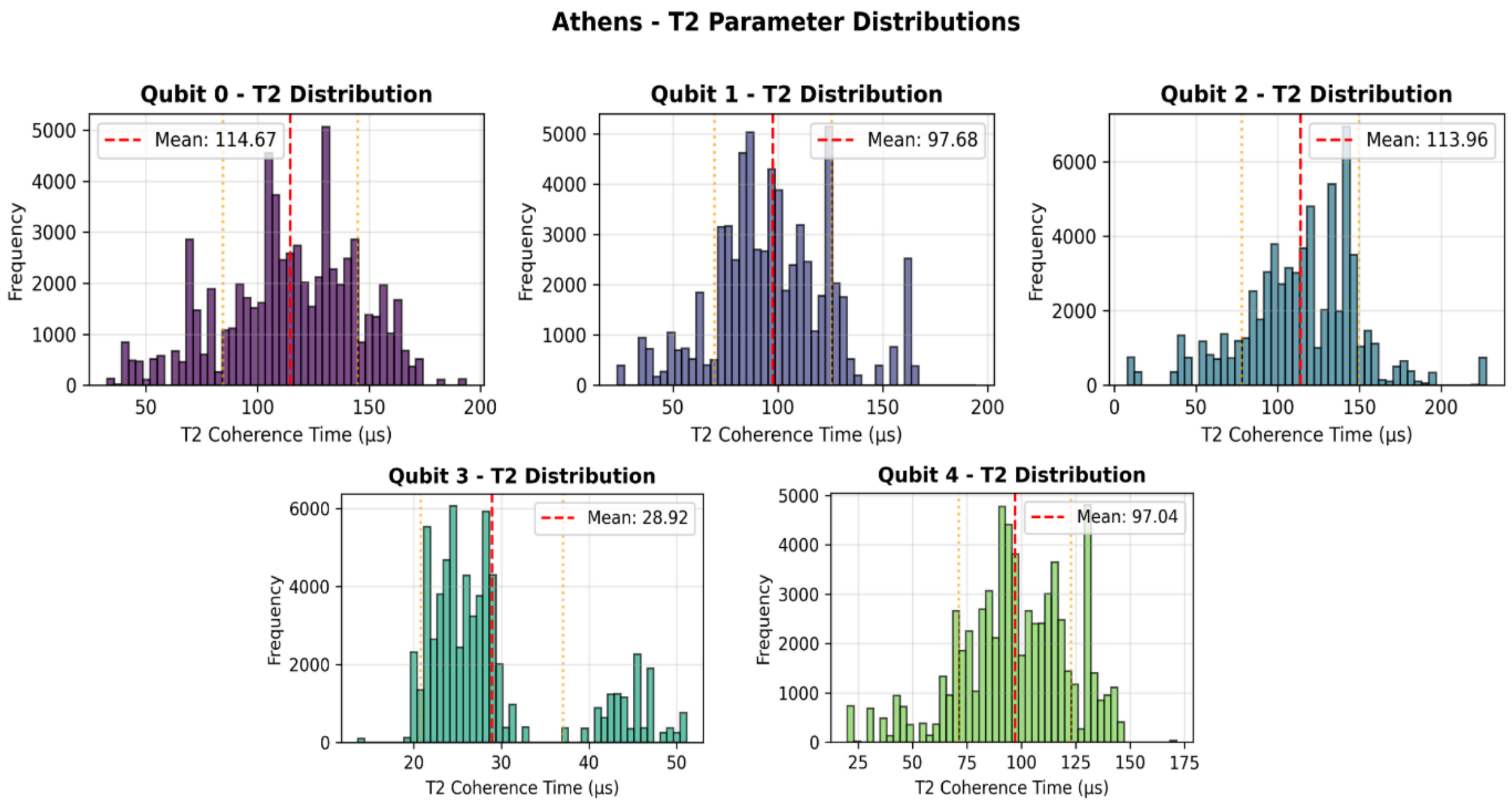

An examination of the range and standard deviation indicates that although Melbourne performs poorly in terms of absolute values, its data show relatively low variability and are more uniform. Conversely, Santiago, despite offering the best averages, displays larger fluctuations across qubits, suggesting that its superior performance may not be uniformly distributed. Athens provides an intermediate balance, combining moderate averages with acceptable stability.

In summary, the Santiago processor is the most suitable choice for implementing and testing QPUF designs, as it provides the best combination of low readout error and long relaxation and dephasing times. However, its larger variability across qubits should be considered when precision consistency is required. Athens represents a balanced alternative, while Melbourne, due to its high readout error and short coherence times, is unsuitable for noise-sensitive quantum applications.

***Figure 6.*** *Frequency distribution of readout error per physical qubit on Athens*

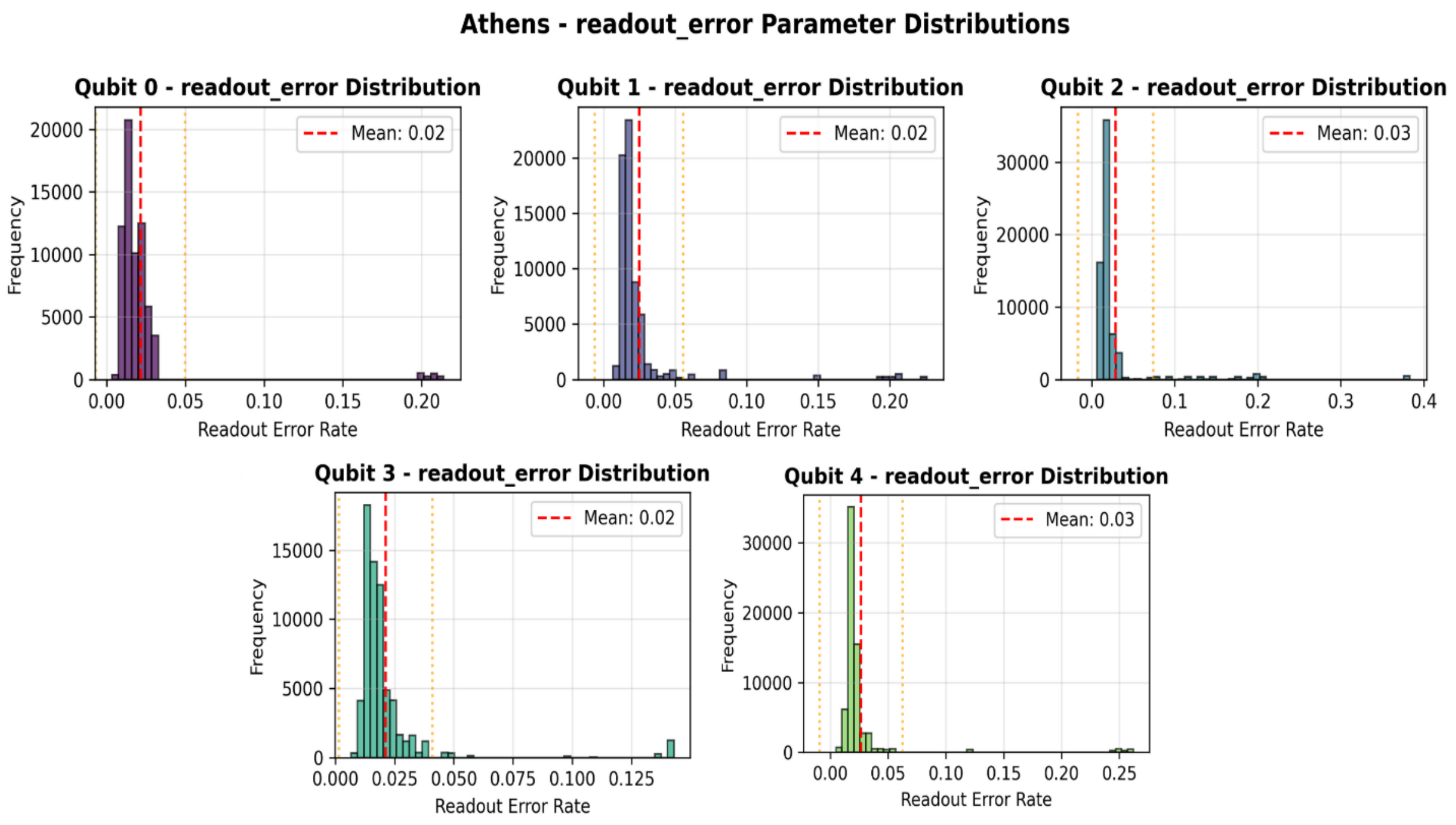

***Figure 7.*** *Frequency distribution of $T_1$ values per physical qubit on Santiago*

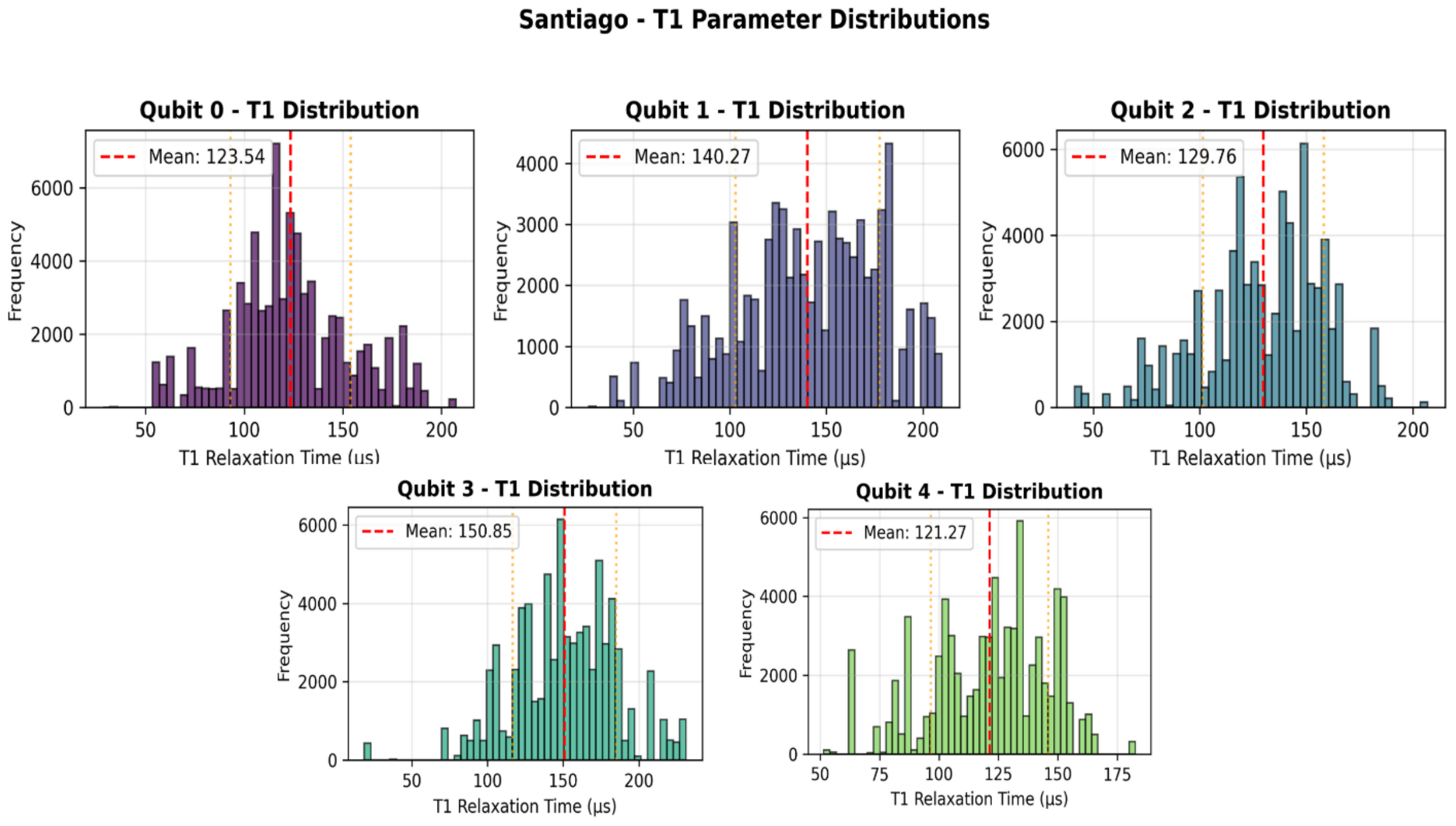

***Figure 8.*** *Frequency distribution of $T_2$ values per physical qubit on Santiago*

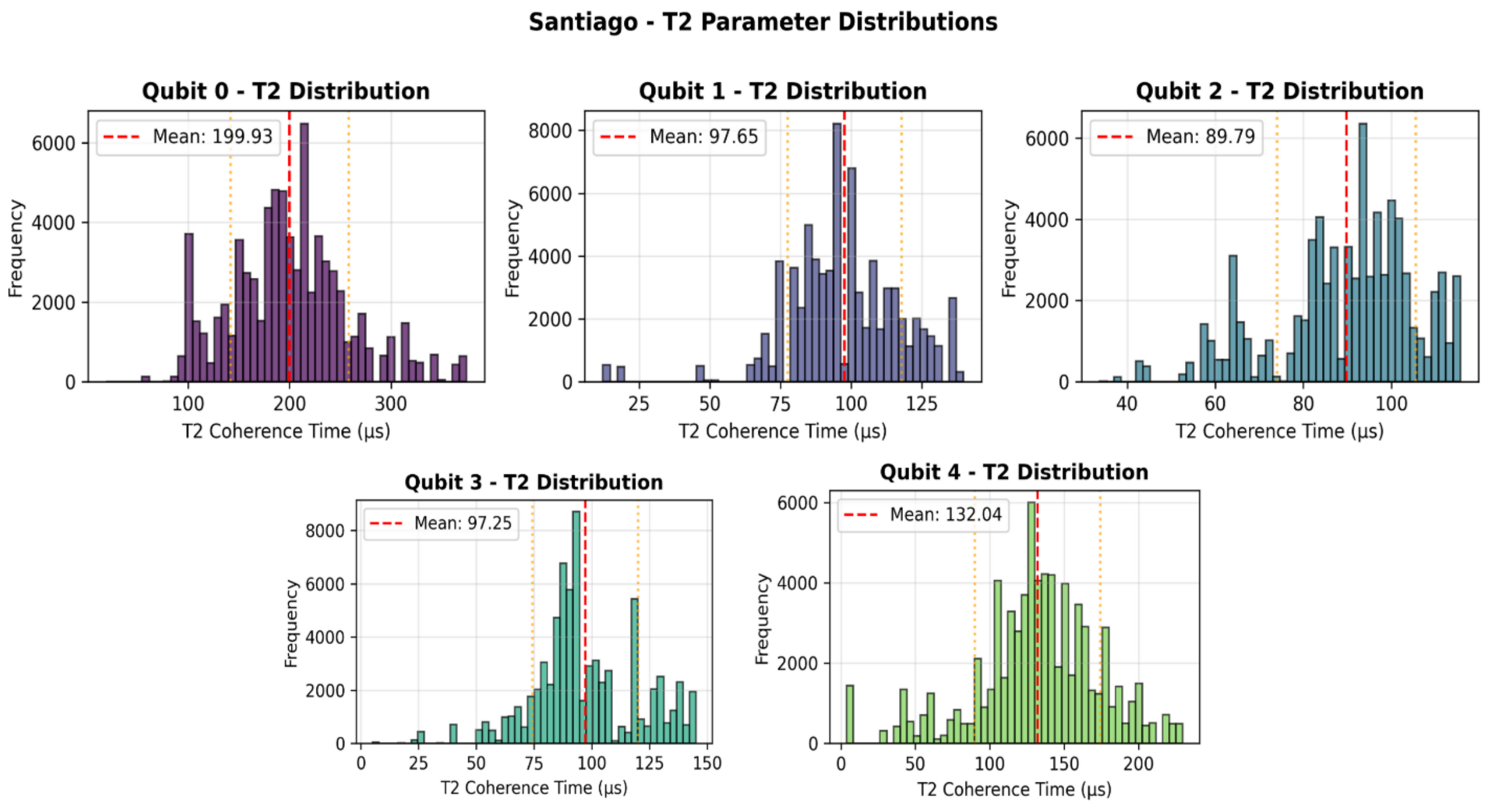

***Figure 9.*** *Frequency distribution of readout error per physical qubit on Santiago*

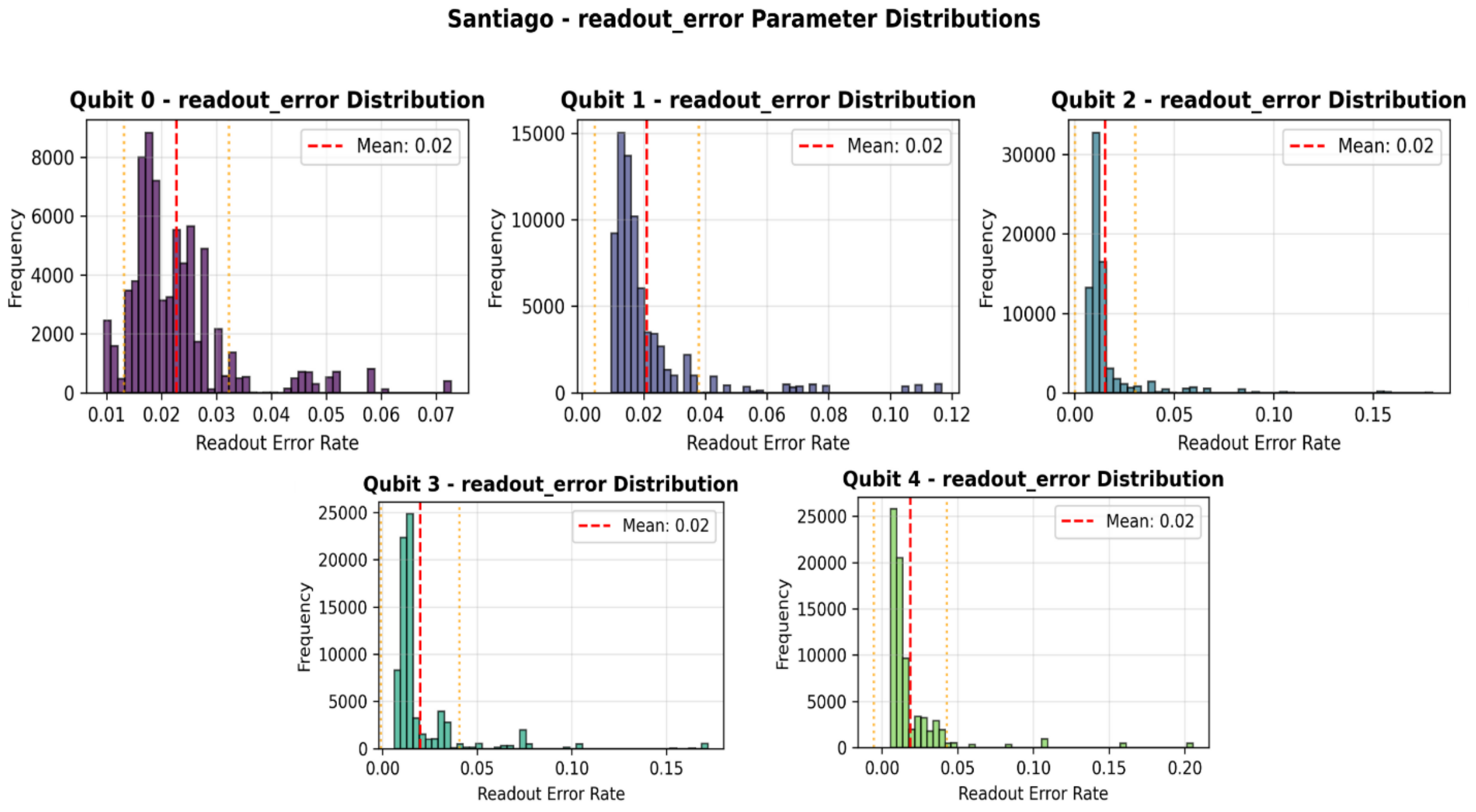

***Figure 10.*** *Frequency distribution of $T_1$ values per physical qubit on Melbourne*

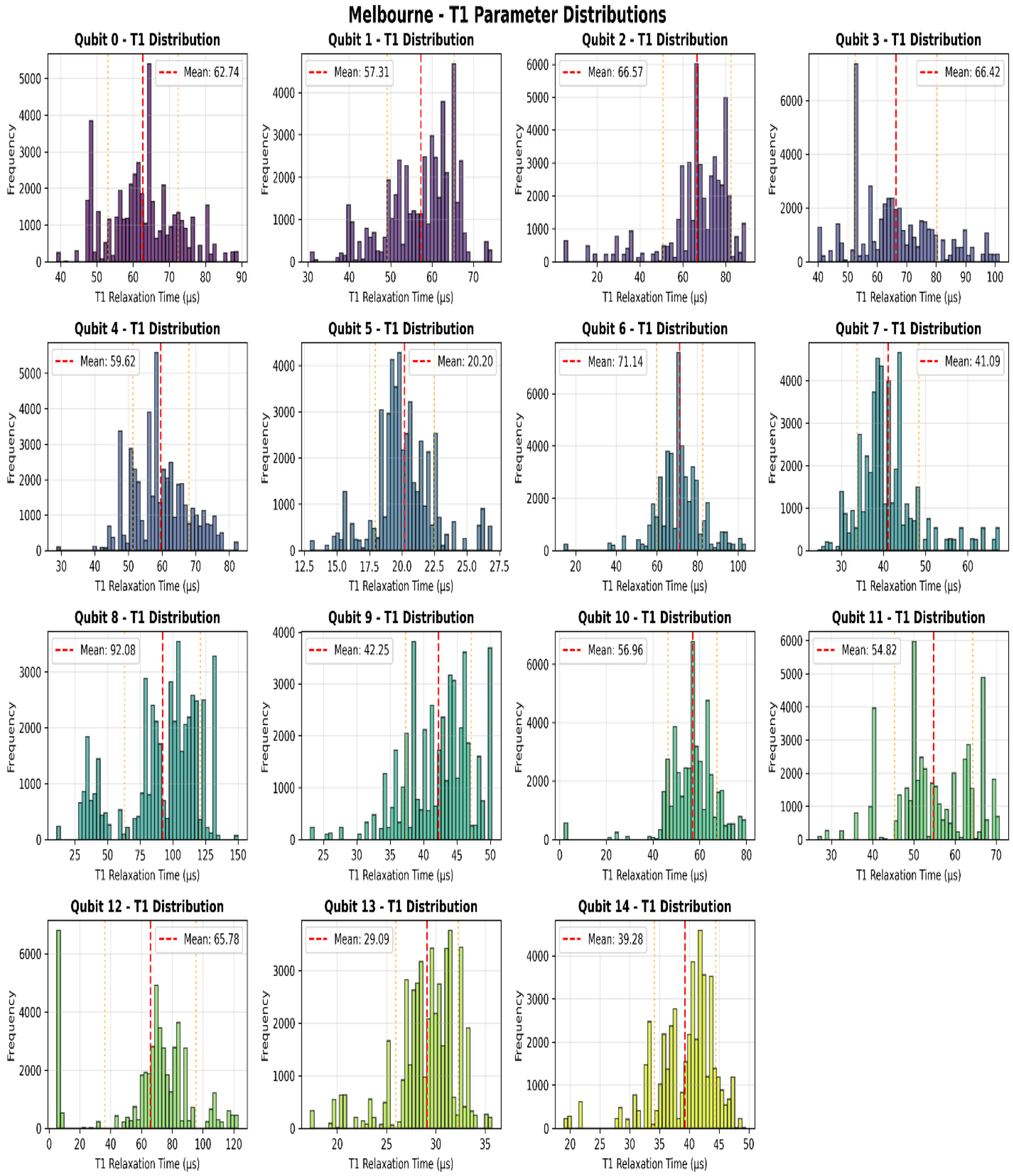
Melbourne - T1 Parameter Distributions
Qubit 0 - T1 Distribution
Mean: 62.74
Qubit 1 - T1 Distribution
Mean: 57.31
Qubit 2 - T1 Distribution
Mean: 66.57
Qubit 3 - T1 Distribution
Mean: 66.42
Qubit 4 - T1 Distribution
Mean: 59.62
Qubit 5 - T1 Distribution
Mean: 20.20
Qubit 6 - T1 Distribution
Mean: 71.14
Qubit 7 - T1 Distribution
Mean: 41.09
Qubit 8 - T1 Distribution
Mean: 92.08
Qubit 9 - T1 Distribution
Mean: 42.25
Qubit 10 - T1 Distribution
Mean: 56.96
Qubit 11 - T1 Distribution
Mean: 54.82
Qubit 12 - T1 Distribution
Mean: 65.78
Qubit 13 - T1 Distribution
Mean: 29.09
Qubit 14 - T1 Distribution
Mean: 39.28
Frequency
T1 Relaxation Time (μs)

***Figure 11.*** *Frequency distribution of $T_2$ values per physical qubit on Melbourne*

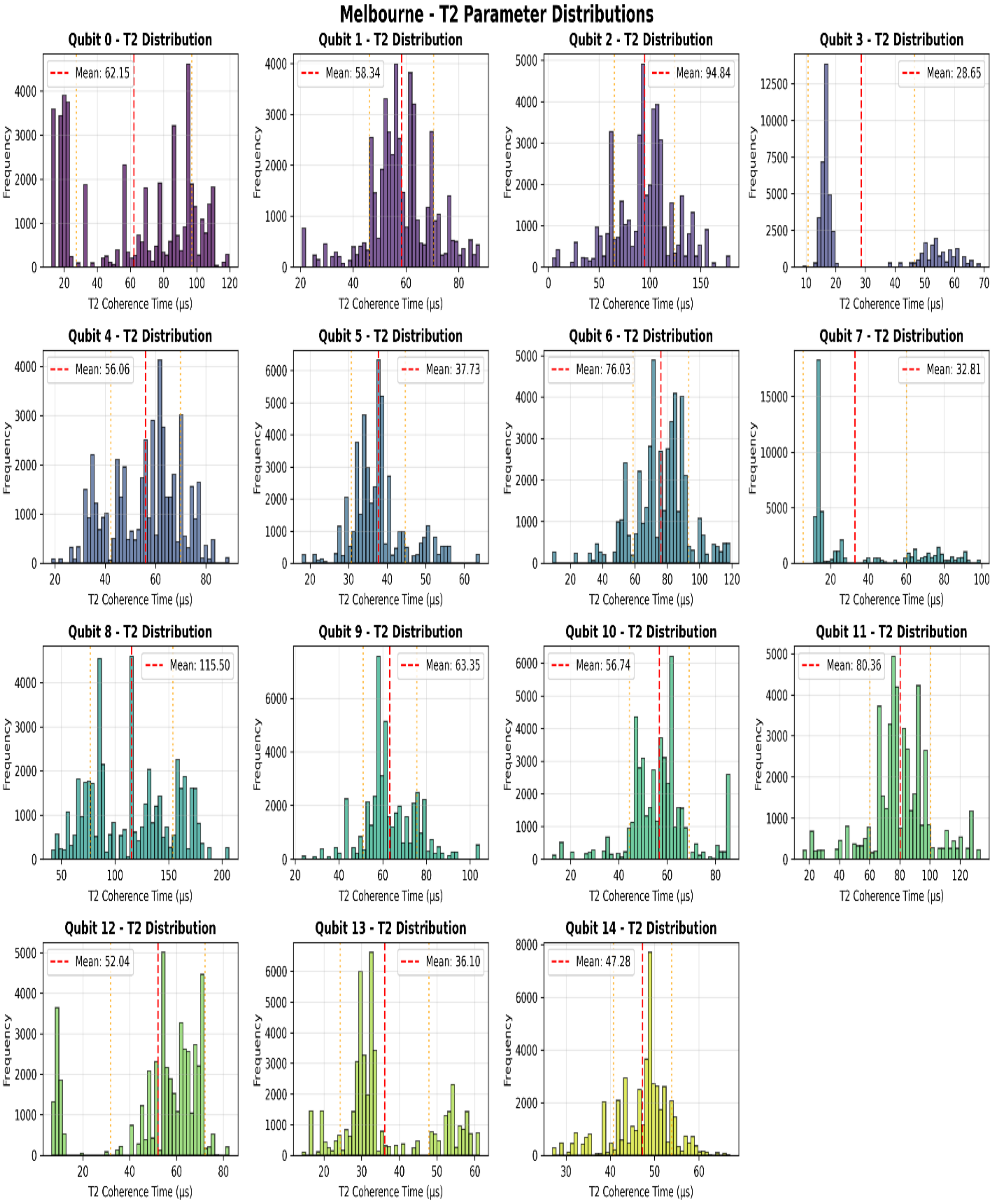

***Figure 12.*** *Frequency distribution of readout error per physical qubit on Melbourne*

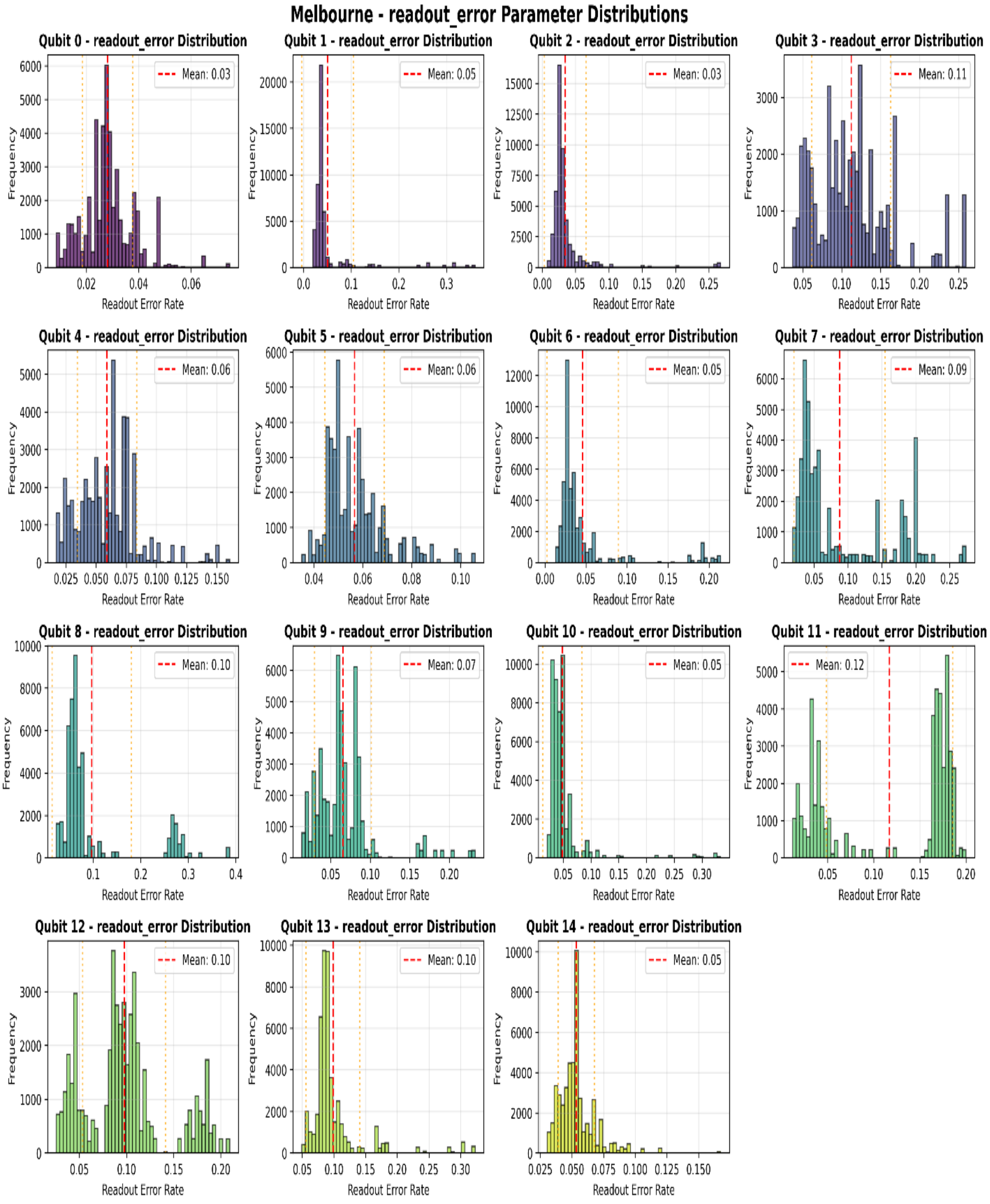

*5.4. D-QPUF simulation results*

This section presents the simulation procedure and results of the Dissipative Quantum Physically Unclonable Function (D-QPUF) design. The simulation was conducted first using the Aer backend to validate the theoretical model under idealized noise conditions, followed by experiments on the Athens and Santiago processors to evaluate real-device behaviour.

To implement the D-QPUF quantum circuit, noise coefficients were generated dynamically based on the input challenge bits for three primary noise channels: decoherence, dephasing, and depolarization. The coefficients were defined as follows:

$$\gamma_{\text{amp}} = 0.01 + 0.005 \times (\text{Number of 1-bits in the challenge})$$
$$\gamma_{\text{phase}} = 0.02 + 0.003 \times (\text{Number of 0-bits in the challenge})$$
$$p_{\text{depol}} = 0.01 + 0.002 \times (\text{Total number of challenge bits})$$

This formulation establishes a direct dependency between the challenge and the noise model. Each unique challenge therefore induces a distinct noisy environment, ensuring that even if two circuits share the same logical structure, they yield different responses due to their noise-induced variability. The baseline values (0.01 and 0.02) were chosen to reflect realistic error rates of superconducting qubits in IBM Quantum devices, while the incremental coefficients were tuned to produce noticeable but controlled noise variation. The higher weighting for decoherence and the moderate increments for phase and depolarization maintain the natural balance observed in physical noise dynamics. Using the total number of bits in the depolarization term introduces a quasi-periodic nonlinear source, further diversifying the response space.

By multiplying the coefficients by the count of 1s and 0s in the challenge, the structure of the challenge directly influences the noise intensity. A challenge with more 1-bits experiences stronger decoherence—since qubits in the excited state are less stable—while challenges with more 0-bits experience stronger phase noise, reflecting the higher susceptibility of ground states to environmental interactions. For depolarization, the total bit count was taken modulo 5 to create a five-state cyclical behaviour, introducing nonlinear diversity without excessive randomness. This design ensures that both the pattern and composition of each challenge affect the system's noise dynamics, enhancing the uniqueness and unpredictability of D-QPUF responses.

The D-QPUF circuit configurations for 2, 4, 6, and 8-qubit implementations are shown in Figures 13–16, respectively. Each configuration employs randomized single-qubit and entangling gates to realize non-reproducible mappings while maintaining scalability. The D-QPUF quantum circuit was composed of the following layers:

- **Layer 1:** Random application of $R_x$, $R_y$ and $R_z$ gates determined by a unique identifier for each device instance. This guarantees intrinsic diversity between different realizations of the QPUF.
- **Layer 2:** Entanglement generation via controlled-$X$ $(CX)$ gates between corresponding qubits, producing mixed states that cannot be described by simple linear mappings.
- **Layer 3:** Conditional application of Hadamard $(H)$, phase $(S)$, or random rotation gates dependent on the qubit index and challenge bits:

$$U_i = \begin{cases} H, & \text{if } i \equiv 1 (\text{mod} 4) \\ S, & \text{if } i \equiv 3 (\text{mod} 4) \end{cases} \text{(odd indices)}$$

$$U_i = \begin{cases} R_z(\theta_i), & \text{if } i \equiv 0 (\text{mod} 4) \\ R_y(\theta_i), & \text{if } i \equiv 2 (\text{mod} 4) \end{cases} \text{(even indices)}$$

- **Layer 4:** Adaptive CX patterns based on the overall parity of the challenge. For an even number of 1-bits, a chain topology (1 → 2 → 3 → 4 ...) is used; for an odd sum, a star topology centered on qubit 1 is applied. This global dependency introduces structural variability in the entanglement pattern.

- **Final Layer:** Application of small random $R_x$, $R_y$ and $R_z$ rotations to each qubit, serving as a post-processing stage to enhance response diversity and minimize overlap among challenge-response pairs.

***Figure 13.*** *D-QPUF quantum circuit using 2 qubits*

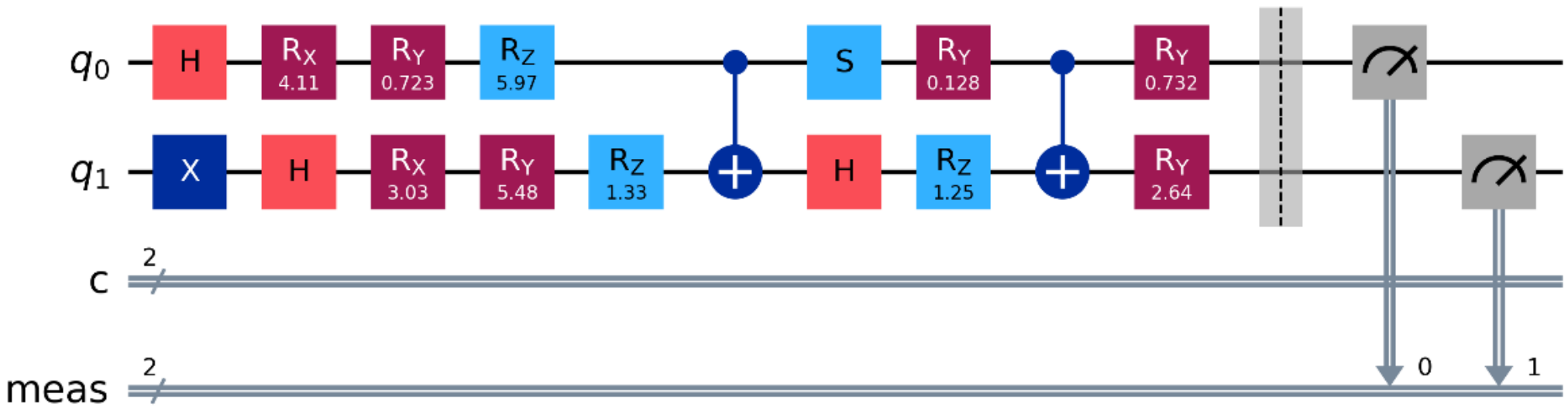

***Figure 14.*** *D-QPUF quantum circuit using 4 qubits*

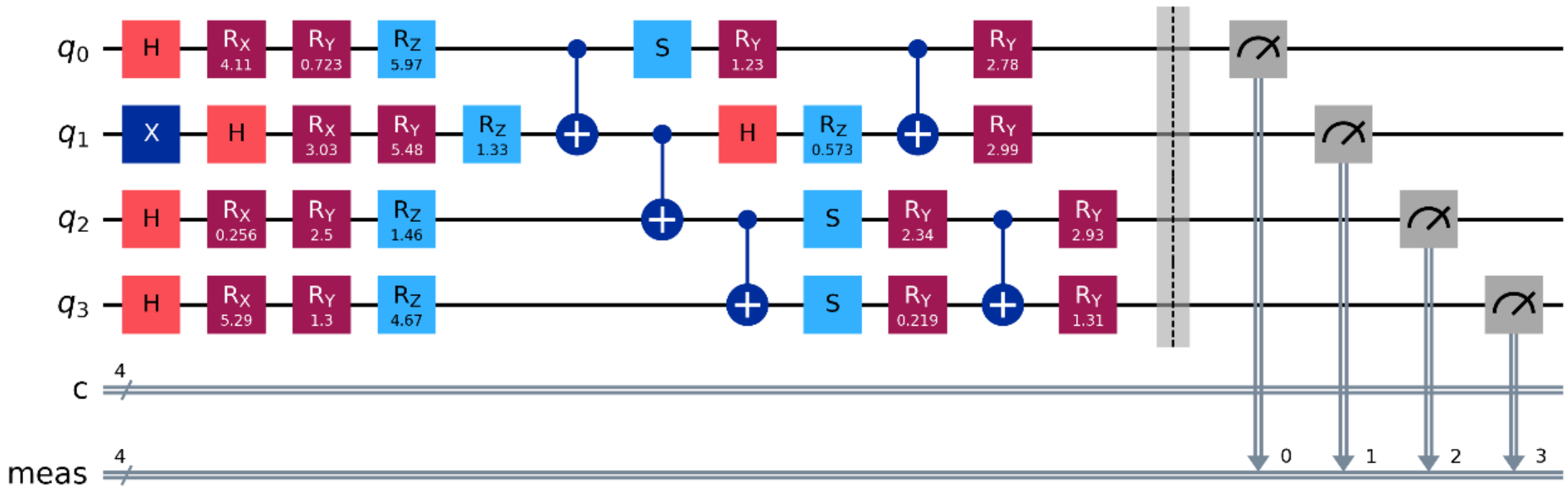

***Figure 15.*** *D-QPUF quantum circuit using 6 qubits*

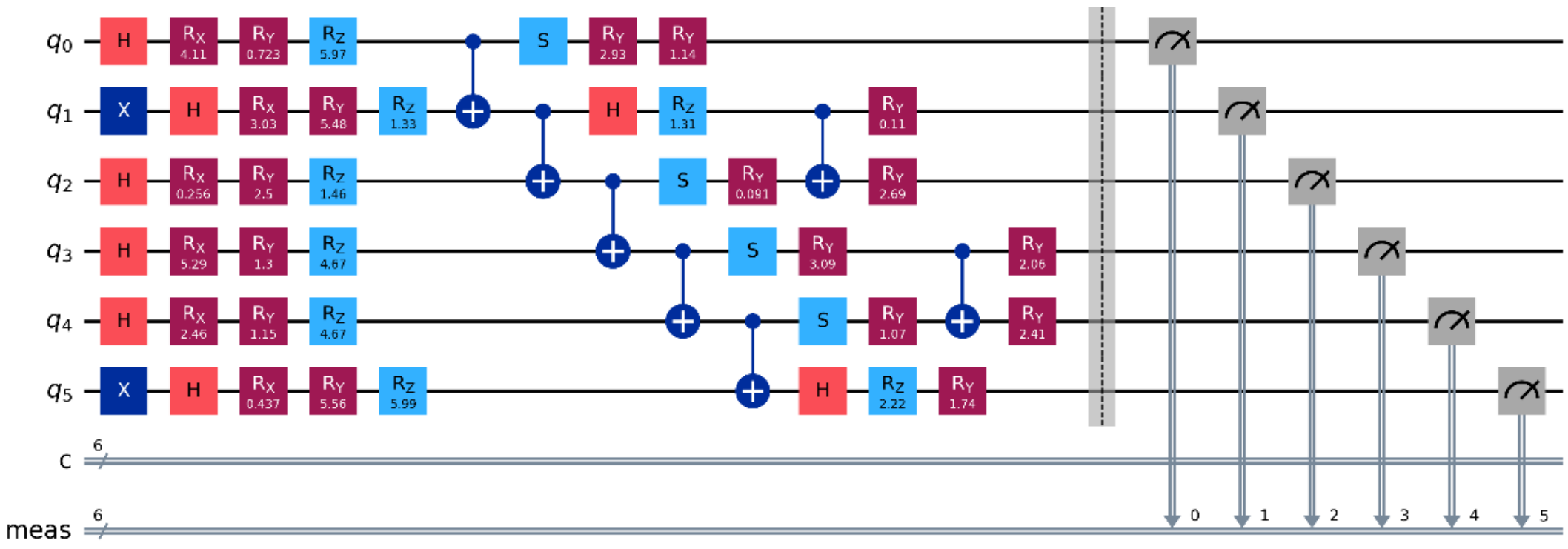

***Figure 16.*** *D-QPUF quantum circuit using 8 qubits*

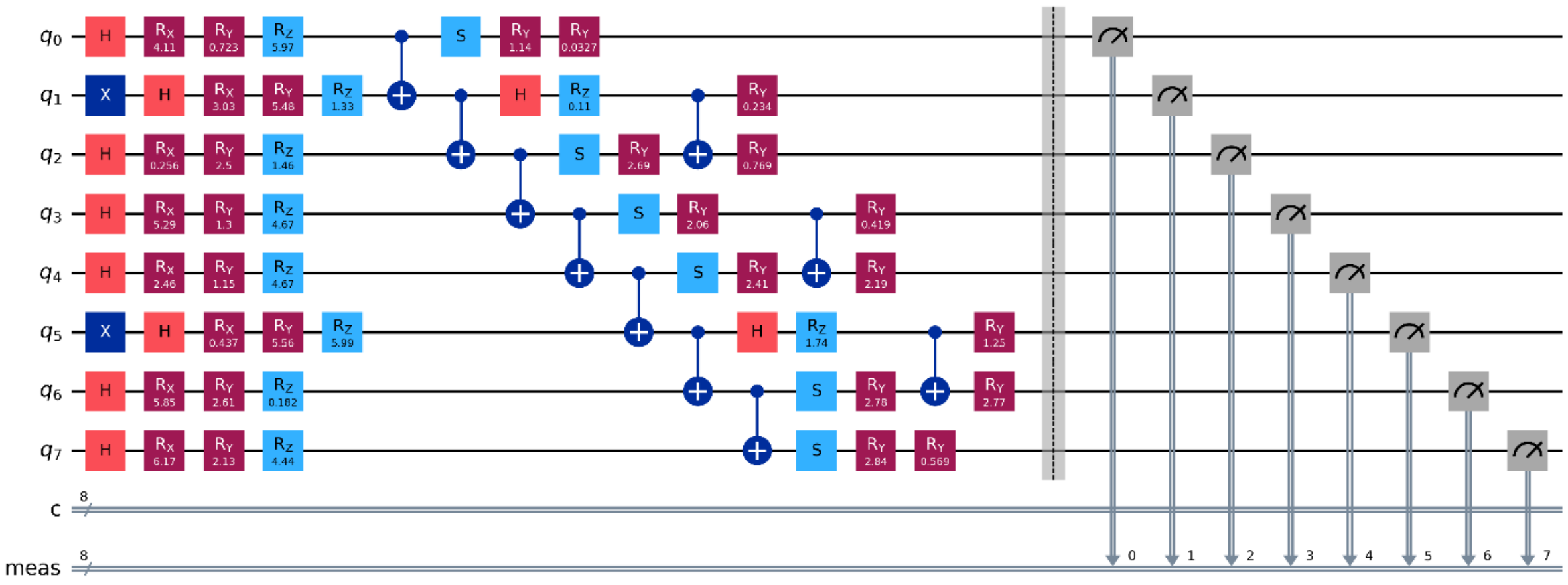

Four D-QPUF circuits (2, 4, 6, and 8-qubit) were simulated on IBM's Aer backend. The results for uniformity, uniqueness, and reliability are presented in Table 4.

***Table 4.*** *Results of Uniformity, Uniqueness, and Reliability for D-QPUF Simulation on the Aer Simulator*

| **Metric/#Qubits** | **2** | **4** | **6** | **8** |
|---|---|---|---|---|
| **Uniformity** | 49.9 | 49.4 | 48.1 | 48.8 |
| **Uniqueness** | 50.6 | 50.1 | 49.9 | 50 |
| **Reliability** | 98.2 | 93 | 87.9 | 77.7 |

The results indicate that the D-QPUF maintains near-ideal uniformity and uniqueness under simulated noise, with only slight deviations as qubit count increases. The best uniformity (49.9%) was observed in the 2-qubit configuration, while the 8-qubit design showed the highest unpredictability due to inter-qubit correlations. Uniqueness values remained consistently close to the ideal 50%, confirming strong challenge sensitivity. Reliability decreased as qubit number increased, primarily due to cumulative gate errors in deeper circuits. Overall, the Aer simulations confirm that the D-QPUF design achieves balanced uniformity and uniqueness, with reliability being the only metric affected by circuit depth.

Subsequent experiments were performed on the Athens and Santiago quantum processors for configurations ranging from 2 to 5 qubits. The results are summarized in Tables 5 and 6.

The comparison reveals clear differences between the two processors. *Athens*, characterized by higher readout error (2.45%) and larger fluctuations in $T_1$ and $T_2$, demonstrates substantial reliability degradation even in small circuits. Its linear topology necessitates additional SWAP operations for entanglement beyond two qubits, compounding gate errors and reducing performance.

Conversely, Santiago—with lower readout error (1.95%) and higher coherence times—exhibits significantly better performance. Its star topology minimizes SWAP operations, thereby preserving uniqueness and achieving higher reliability. Although reliability still decreases with circuit size, the reduction is less severe than on Athens, confirming the positive effect of improved coherence and topology.

***Table 5.*** *Results of Uniformity, Uniqueness, and Reliability for D-QPUF Simulation on Athens*

| Metric/#Qubits | 2 | 3 | 4 | 5 |
|---|---|---|---|---|
| **Uniformity** | 29.71 | 29.5 | 30.12 | 28.47 |
| **Uniqueness** | 41.71 | 40.76 | 41.35 | 40.43 |
| **Reliability** | 71.2 | 64.8 | 60.6 | 59.8 |

***Table 6.*** *Results of Uniformity, Uniqueness, and Reliability for D-QPUF Simulation on Santiago*

| Metric/#Qubits | 2 | 3 | 4 | 5 |
|---|---|---|---|---|
| **Uniformity** | 57.32 | 55.74 | 55.62 | 54.41 |
| **Uniqueness** | 48.3 | 48.44 | 48.58 | 48.94 |
| **Reliability** | 91 | 85.6 | 79.4 | 70.4 |

These findings indicate that the magnitude and variability of $T_1$ and $T_2$, coupled with topological constraints, directly determine the effectiveness of noise-injection-based QPUF designs. Consequently, D-QPUFs do not yield uniform performance across different quantum computers. Nonetheless, the consistent uniformity trends and the isolated reliability decline on Santiago suggest that performance could be further improved through stability-enhancing calibration or adaptive error mitigation techniques.

It is worth noting that ongoing advancements in quantum hardware—particularly the reduction of $T_1$ and $T_2$ fluctuations—will likely enable even higher stability in future D-QPUF implementations. Reassessing the proposed design on newer processors thus represents a promising avenue for validation and optimization.

### *5.5. MF-QPUF simulation results*

The measurement-feedback QPUF (MF-QPUF) extends the D-QPUF architecture by introducing mid-circuit measurements whose classical outcomes are immediately fed back to condition subsequent unitary operations. The implementation follows the same layered approach used for D-QPUF, but with a crucial modification in the third layer: certain qubits are measured and their outcomes determine which gates are applied to the remaining qubits. Concretely, after the initial state preparation and entangling layers, a subset of qubits is measured in the computational basis; if the measurement of qubit $i$ yields a 1, the feedback logic applies either a Hadamard gate or a $R_z(\theta)$ rotation to a target qubit (predominantly odd-indexed qubits), whereas a 0 result triggers a phase gate $S$ or a small $R_x(\theta)$ or $R_y(\theta)$ rotation on mainly even-indexed targets. This mid-circuit decision makes the evolution adaptive and intrinsically stochastic: the final channel is the mixture of branches associated with all possible measurement outcomes and corresponding conditional unitaries.

Simulations of MF-QPUF were first performed on the noise-free Aer simulator to characterize the idealized behaviour of this feedback-driven design. Implementations for 2, 4, 6, and 8 qubits were examined, and the standard QPUF quality metrics—uniformity, uniqueness, and reliability—were recorded. The measured uniformity, uniqueness, and reliability values for the MF-QPUF architecture are listed in Tables 7 and 8, corresponding to simulations on the Aer backend and experimental runs on the Santiago device. The results confirm the robustness of the measurement-feedback mechanism across different hardware environments. Figures 17–20 depict the MF-QPUF architectures with 2, 4, 6, and 8 qubits. In these circuits, mid-circuit measurements are used to generate classical feedback that conditions subsequent unitary operations, introducing controlled stochasticity in the challenge–response behaviour. On Aer the MF-QPUF exhibits a nontrivial dependence of metrics on system size. Uniformity and uniqueness peak near intermediate sizes: the 4-qubit configuration achieved the best balance with uniformity ≈ 46.1% and uniqueness ≈ 45.6%, while the 2-qubit circuit showed the poorest statistical balance (uniformity ≈ 35.5%, uniqueness ≈ 36.8%). Reliability on Aer declines with increasing qubit count: the 2-qubit circuit achieves the highest reproducibility (≈ 63.5%), but the 8-qubit circuit drops to ≈ 40.1%. These trends reflect the tradeoffs intrinsic to measurement-feedback designs: a single feedback round ($f = 1$) is sufficient to introduce substantial branch diversity, improving output entropy at moderate sizes, but as the circuit grows the compounded effect of branching and increased two-qubit interactions amplifies instabilities and reduces repeatability.

***Figure 17.*** *MF-QPUF quantum circuit using 2 qubits*

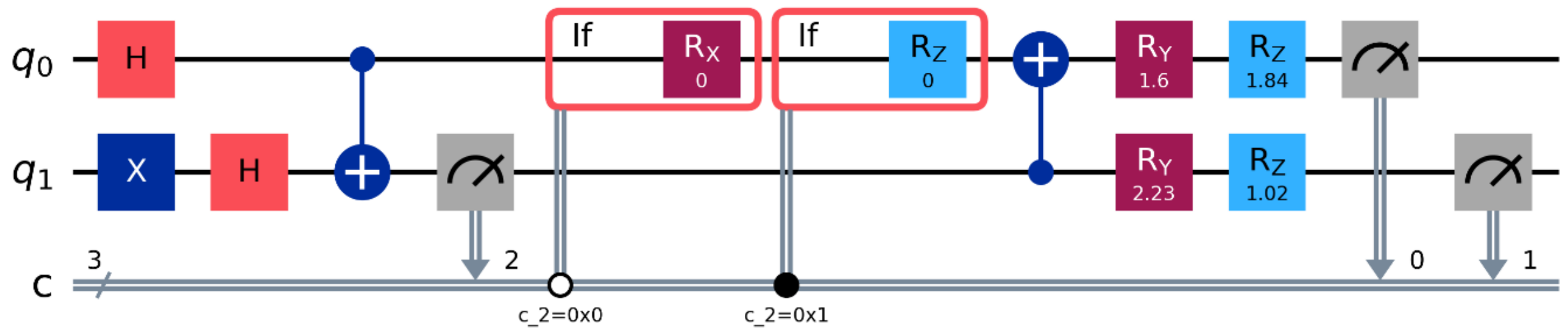

***Figure 18.*** *MF-QPUF quantum circuit using 4 qubits*

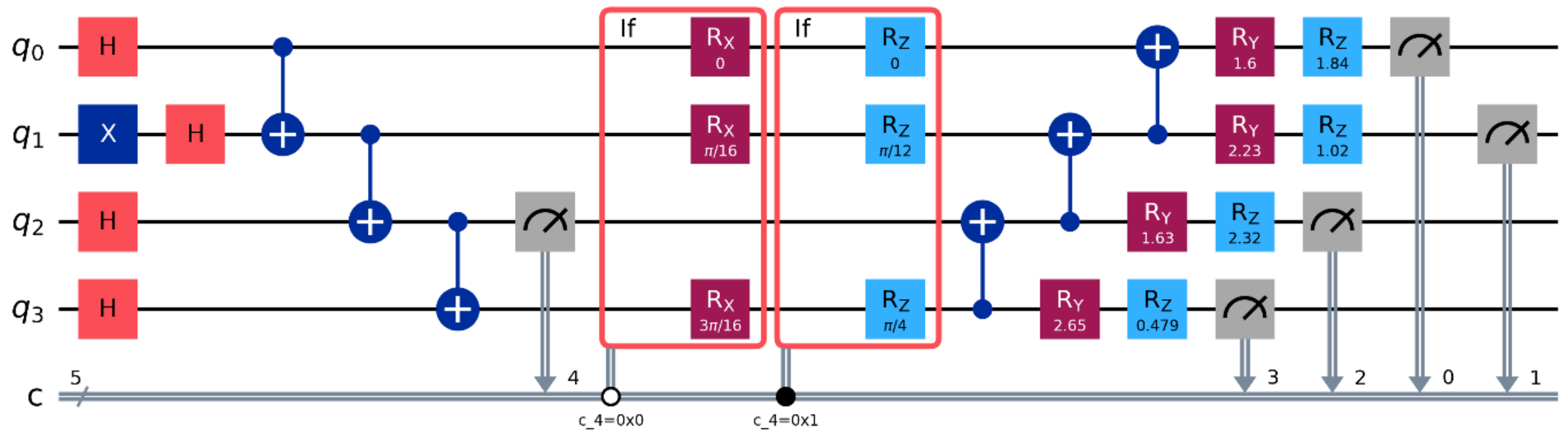

***Figure 19.*** *MF-QPUF quantum circuit using 6 qubits*

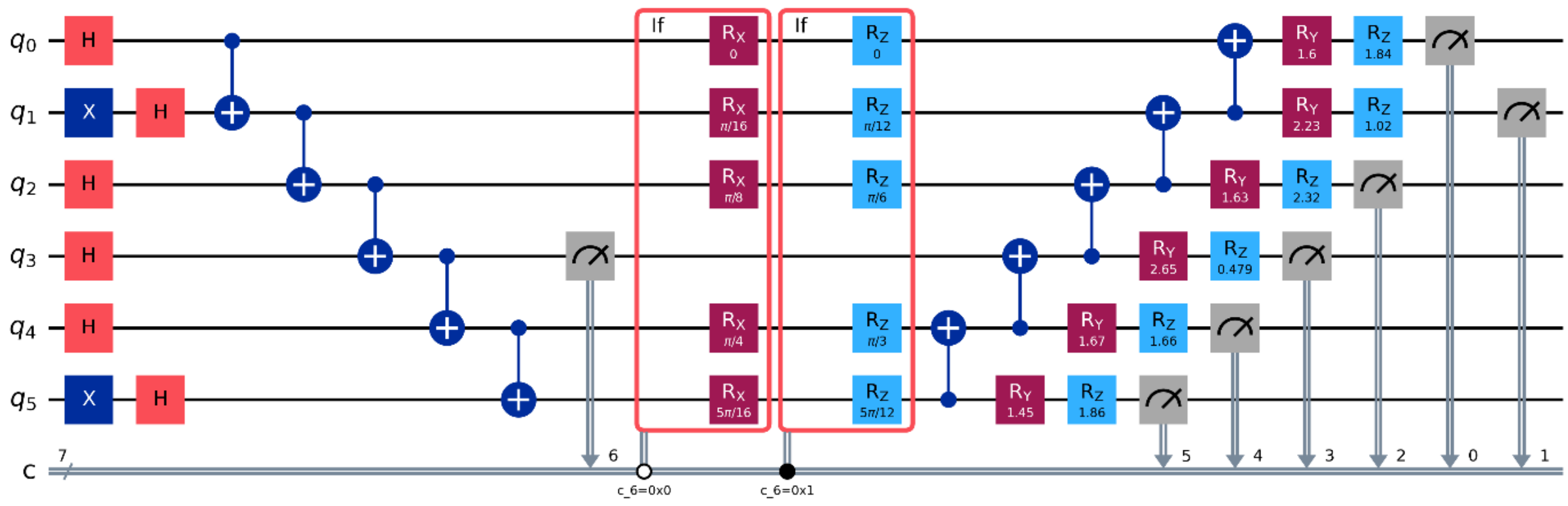

***Figure 20.*** *MF-QPUF quantum circuit using 8 qubits*

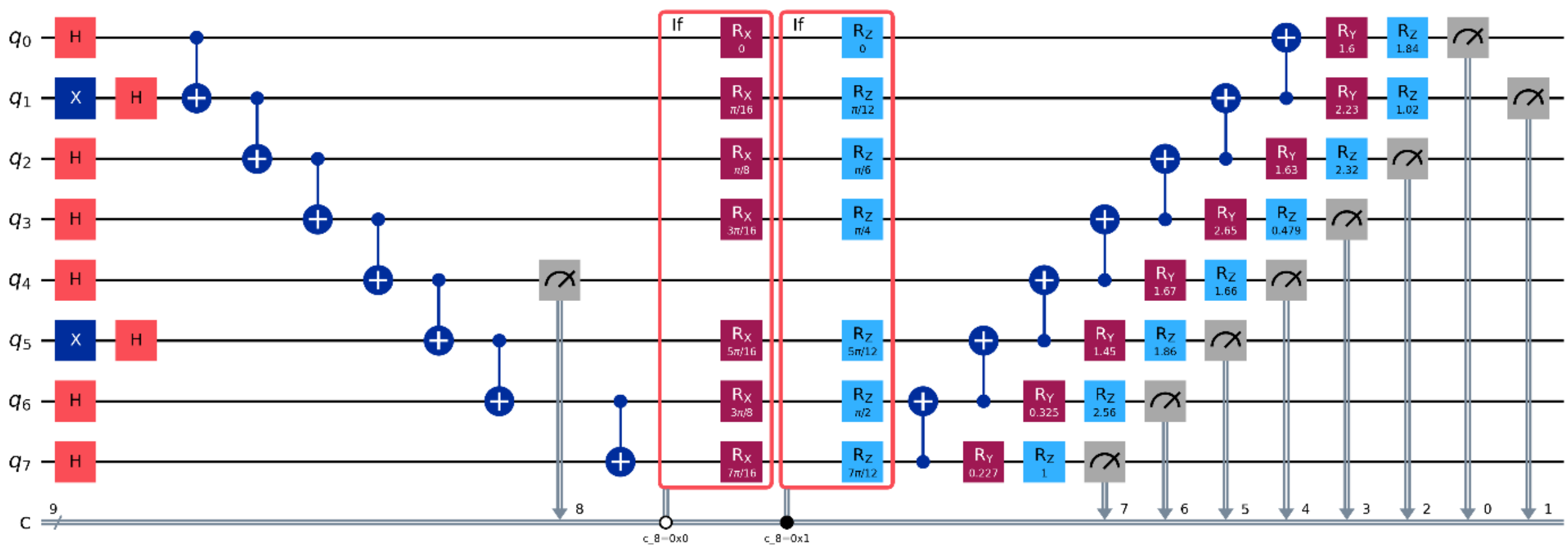

To assess real-device performance and topology effects, MF-QPUF was executed on the Santiago backend. The hardware results follow the same qualitative pattern as the Aer simulations but with differences attributable to finite coherence and readout fidelity. On Santiago the 2-qubit MF-QPUF improves in statistical balance compared to Aer: uniformity rises from 35.5% to 39.4% and uniqueness from 36.8% to 38.9%, while reliability decreases slightly (63.5% → 60.7%). At intermediate sizes the hardware impact becomes more pronounced. The 3-qubit Santiago run shows near-ideal statistical quality (uniformity ≈ 48.7%, uniqueness ≈ 47.2%), but reliability is only ≈ 55.2%. The 4-qubit case experiences a sharp reliability drop on Santiago (57.1% on Aer → 39.0% on Santiago) despite a minor change in uniformity and uniqueness (46.1% → 44.4% and 45.6% → 45.9%, respectively). For 5-qubit circuits the hardware overhead of routing and additional two-qubit gates further degrades reproducibility (reliability ≈ 34.1%), although uniformity and uniqueness remain in acceptable ranges (≈ 42.4% and ≈ 44.2%).

These observations can be interpreted from both operational and physical viewpoints. Mid-circuit measurements and feedforward create many computational branches; each branch is a different effective circuit whose success depends on qubit coherence during the measurement–feedback latency and on readout fidelity. Thus, while the feedback mechanism amplifies randomness and can improve uniformity and uniqueness by diversifying outputs, it also increases vulnerability to readout errors, idle decoherence, and SWAP overheads caused by topology mapping. Santiago's relatively long $T_1$ and $T_2$ and modest readout error (≈ 2%) help preserve statistical balance compared with Aer's idealized model, but the physical cost of conditional operations and longer circuits manifests as a clear decline in reliability as qubit number grows.

***Table 7.*** *Results of Uniformity, Uniqueness, and Reliability for MF-QPUF Simulation on the Aer Simulator*

| Metric/#Qubits | 2 | 4 | 6 | 8 |
|---|---|---|---|---|
| **Uniformity** | 35.5 | 46.1 | 44.5 | 41.7 |
| **Uniqueness** | 36.8 | 45.6 | 43.8 | 39.9 |
| **Reliability** | 63.5 | 57.1 | 40.3 | 40.1 |

***Table 8.*** *Results of Uniformity, Uniqueness, and Reliability for MF-QPUF Simulation on Santiago*

| Metric/#Qubits | 2 | 3 | 4 | 5 |
|---|---|---|---|---|
| **Uniformity** | 39.4 | 48.7 | 44.4 | 42.4 |
| **Uniqueness** | 38.9 | 47.2 | 45.9 | 44.2 |
| **Reliability** | 60.7 | 55.2 | 39 | 34.1 |

Comparing MF-QPUF to D-QPUF reveals the expected trade-off: MF-QPUF attains higher statistical diversity (in some cases superior uniformity and uniqueness) at the cost of lower reproducibility. D-QPUF avoids mid-circuit collapse and therefore sustains higher reliability, while MF-QPUF leverages internal randomness to increase the entropy of responses. In practice, the choice between these designs depends on application requirements: MF-QPUF is preferable when challenge diversity and unforgeability are paramount and moderate unreliability is tolerable, whereas D-QPUF suits scenarios demanding strong repeatability.

Overall, the MF-QPUF experiments demonstrate that measurement-feedback mechanisms are an effective route to increase the expressive power of QPUFs, but they impose stringent hardware requirements (fast, high-fidelity mid-circuit readout and low latency feedforward) to maintain acceptable reliability. On current NISQ backends, MF-QPUF yields promising statistical metrics at small and intermediate scales but fails the reliability criterion for larger circuits without additional error-mitigation or hardware improvements.

### *5.6. L-QPUF simulation results*

The Lindbladian Quantum Physical Unclonable Function (L-QPUF) aims to represent the authentic dynamics of an open quantum system governed by the Lindblad master equation, thereby embedding Markovian noise

processes—such as decay, dephasing, and collective dissipation—directly into the structure of the PUF. Each simulation begins by mapping the classical challenge vector, a binary string of length $n$, onto the initial quantum register. A bit value of 1 triggers the application of an $X$ gate to the corresponding qubit, while 0 leaves it in the ground state $|0\rangle$. This transformation is unitary and reversible, ensuring that the challenge encoding does not introduce premature state collapse. Subsequently, Hadamard gates are applied to all qubits, producing uniform superpositions $|+\rangle^{\otimes n}$ that define the starting state for the system's noisy evolution. To create non-trivial correlations, controlled-NOT gates are inserted between neighbouring qubits, establishing an entangling layer that permits collective decoherence effects to emerge naturally during the subsequent open-system evolution.

The central stage of the L-QPUF simulation reformulates the quantum state as a density matrix $\rho$ and evolves it under the Lindblad equation

$$\dot{\rho} = -i[H_{\text{sys}}, \rho] + \sum_{\alpha} \left( L_\alpha \rho L_\alpha^\dagger - \frac{1}{2}\{L_\alpha^\dagger L_\alpha, \rho\} \right) ,$$

where $H_{\text{sys}}$ is a weak system Hamiltonian composed primarily of local $X_i$ operators, and the set of collapse operators $\{L_\alpha\}$ includes single-qubit amplitude-damping terms, phase-damping (dephasing) channels, and one or more two-qubit collective noise operators. This framework enforces a completely positive, trace-preserving (CPTP) Markovian evolution that depends only on the current system state, with no memory of previous configurations. Because non-unitary time evolution cannot be directly executed on present-day quantum processors, the final state $\rho_{\text{final}}$ is approximated by an equivalent unitary circuit using a Trotter–Suzuki decomposition of the Lindbladian propagator $e^{\mathcal{L}t}$ into short-time segments. Each local block is then expressed in terms of standard quantum gates (full derivations are provided in Appendix A). The expectation values $\text{Tr}(\rho_{\text{final}} H_{\text{sys}})$ and the reduced single-qubit density matrices are mapped to the rotation parameters of $R_y$ and $R_z$ gates, while two-qubit gates such as $CZ$ encode residual non-local correlations. A concluding random-rotation layer consisting of $R_x$, $R_y$, and $R_z$ operations increases entropy and prevents formation of deterministic output patterns. After final measurement of all qubits, the QPUF response is extracted via a majority-vote rule.

Simulations were first carried out on the Aer backend for 2, 4, 6, and 8-qubit configurations. The experimental and simulated results for the L-QPUF are summarized in Tables 9–12. These tables compare performance across Aer, Santiago, and Melbourne devices, demonstrating the stability of the Lindbladian noise-based approach under varying decoherence and readout conditions. The corresponding Lindblad-based circuit layouts for 2, 4, 6, and 8-qubit configurations are presented in Figures 21-24. These designs embed non-unitary quantum channels within the circuit to model dissipative and decoherence processes explicitly, forming the basis for the L-QPUF implementation. In contrast to D-QPUF, where non-unitarity was artificially emulated by gate-level noise, and MF-QPUF, where mid-circuit measurements introduced stochastic feedback, the L-QPUF integrates realistic decoherence processes into its mathematical model. Consequently, its output distributions exhibit superior balance and reproducibility. Across all tested circuit sizes, uniformity on Aer remained between 49.5 % and 53.2 %, extremely close to the ideal 50 %. This stability persisted under hardware noise: simulations on Athens, Santiago, and Melbourne deviated by less than ±0.3 % from the ideal uniform distribution. Such robustness reflects the design principle of L-QPUF—noise is not an external disturbance but an intrinsic part of the system's CPTP dynamics, ensuring that statistical balance is preserved regardless of circuit depth or device noise level.

The uniqueness metric further corroborates this behaviour. On all platforms, uniqueness remained confined to the narrow interval 49.8–50.2 %, nearly ideal and significantly more stable than the fluctuations observed in D-QPUF and MF-QPUF. This outcome arises from the action of the Lindblad jump operators, which generate distinct yet statistically balanced challenge–response pairs by distributing decoherence effects evenly across the Hilbert space. The result is a PUF whose responses are both diverse and unbiased, even under varying environmental conditions.

***Figure 21.*** *L-QPUF quantum circuit using 2 qubits*

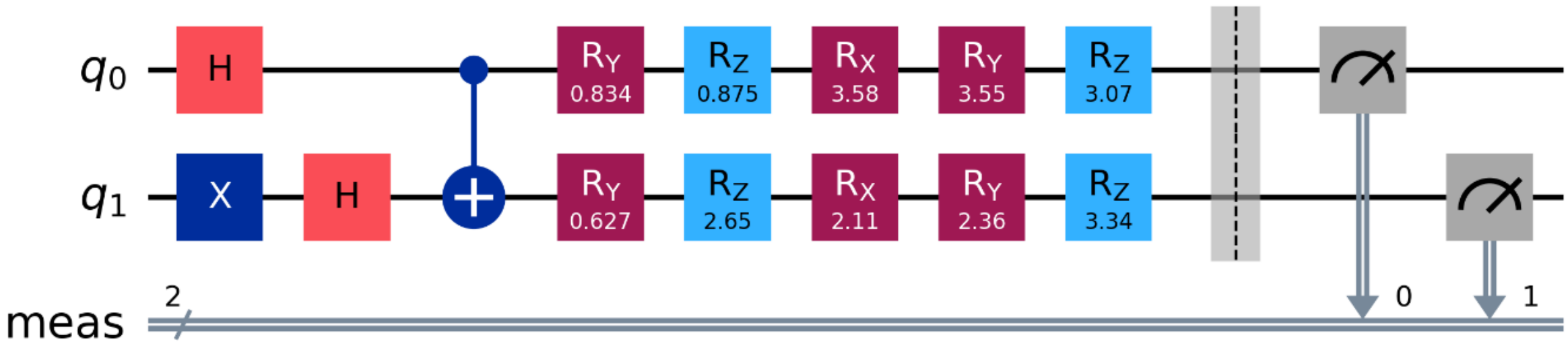

***Figure 22.*** *L-QPUF quantum circuit using 4 qubits*

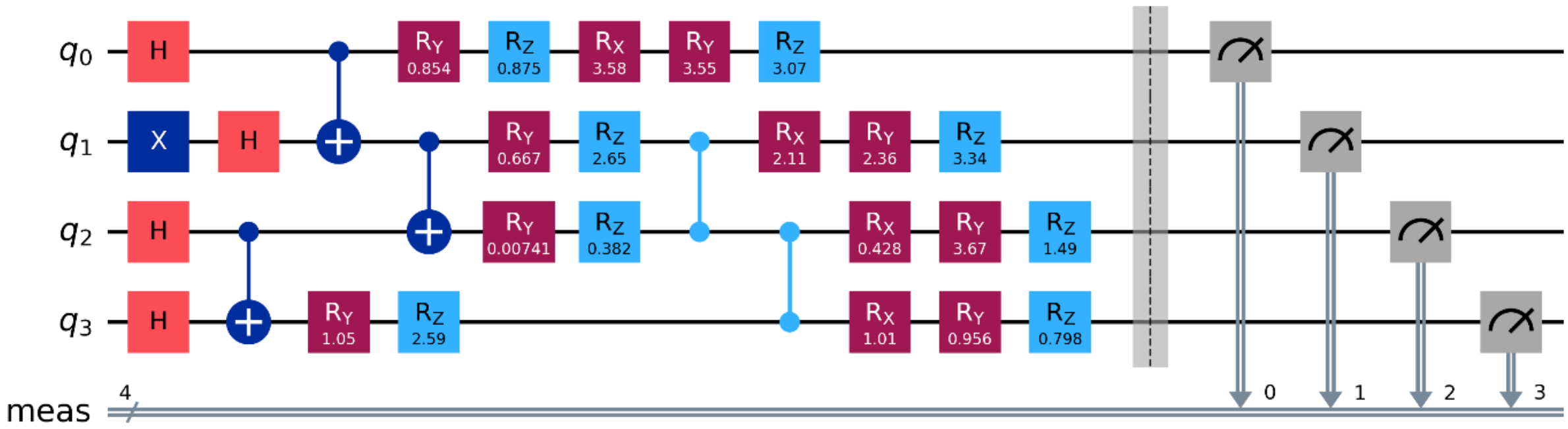

***Figure 23.*** *L-QPUF quantum circuit using 6 qubits*

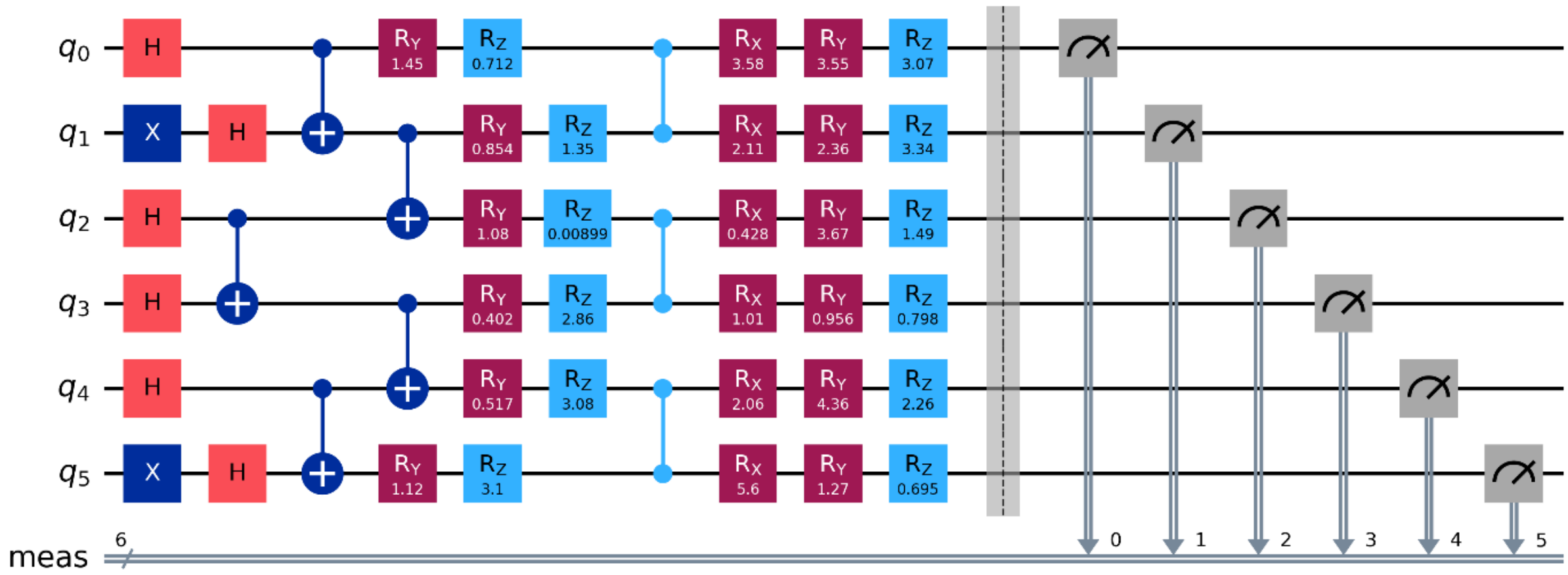

***Figure 24.*** *L-QPUF quantum circuit using 8 qubits*

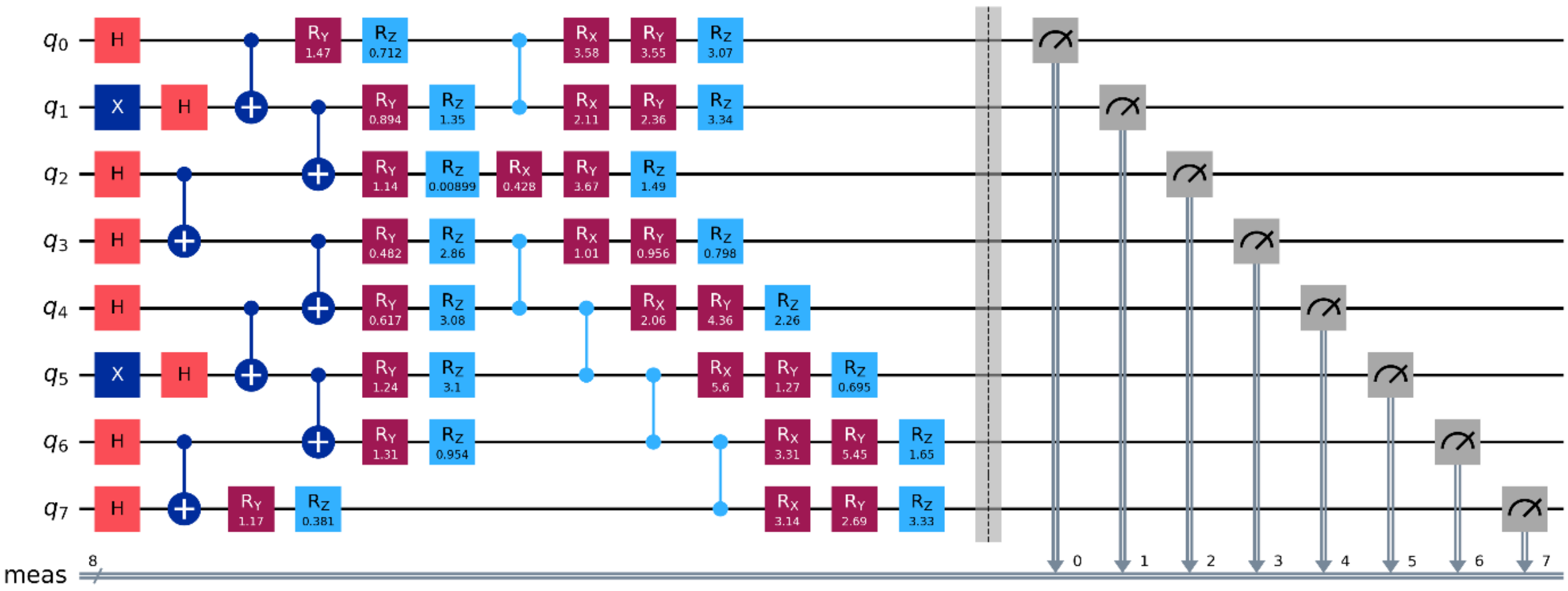

***Table 9.*** *Results of Uniformity, Uniqueness, and Reliability for L-QPUF Simulation on the Aer simulator*

| **Metric/#Qubits** | **2** | **4** | **6** | **8** |
|---|---|---|---|---|
| **Uniformity** | 53.2 | 50 | 49.5 | 50.7 |
| **Uniqueness** | 50.2 | 50.2 | 50.2 | 50.8 |
| **Reliability** | 99.5 | 98.2 | 96.4 | 94.2 |

***Table 10.*** *Results of Uniformity, Uniqueness, and Reliability for L-QPUF Simulation on Athens*

| **Metric/#Qubits** | **2** | **3** | **4** | **5** |
|---|---|---|---|---|
| **Uniformity** | 50.14 | 50.08 | 50.32 | 50.17 |
| **Uniqueness** | 50.03 | 50.16 | 50.1 | 49.99 |
| **Reliability** | 98.44 | 96.97 | 95.41 | 93.08 |

*Table 11. Results of Uniformity, Uniqueness, and Reliability for L-QPUF Simulation on Santiago*

| Metric/#Qubits | 2 | 3 | 4 | 5 |
|---|---|---|---|---|
| **Uniformity** | 50.3 | 50.11 | 50.35 | 50.07 |
| **Uniqueness** | 50.02 | 50.15 | 50.08 | 50.01 |
| **Reliability** | 98.29 | 97.05 | 95.51 | 93.07 |

*Table 12. Results of Uniformity, Uniqueness, and Reliability for L-QPUF Simulation on Melbourne*

| Metric/#Qubits | 5 | 6 | 7 | 8 |
|---|---|---|---|---|
| **Uniformity** | 50.13 | 49.97 | 49.71 | 50.07 |
| **Uniqueness** | 49.97 | 49.98 | 49.89 | 50.01 |
| **Reliability** | 92.05 | 92.73 | 90.84 | 90.14 |

Reliability—the most stringent QPUF metric—shows the most pronounced improvement in L-QPUF. On the Aer simulator, reliability ranged from 94 % to 99.5 %; on Athens and Santiago, it remained between 93 % and 98.5 %; and even on the noisier Melbourne processor, with its relatively short $T_1$ and $T_2$ times and higher readout error, reliability exceeded 90 % in all configurations. The physical explanation lies in the inherent non-unitarity of the Lindblad equation: by embedding environmental interactions directly into the system's formalism, part of the "noise" is reinterpreted as legitimate evolution rather than an external fault. Consequently, repeated evaluations of the same challenge yield consistent responses even when executed on imperfect hardware.

Cross-backend comparison reveals that Athens and Santiago produce nearly indistinguishable results despite their differing $T_1$ and $T_2$ values, demonstrating that L-QPUF maintains stability across devices of varying coherence quality. Melbourne, although limited by higher intrinsic error rates, still preserves strong uniformity, balanced uniqueness, and high reliability—highlighting the architecture's resilience on legacy or mid-fidelity platforms. The principal limitation of L-QPUF is computational rather than physical: density-matrix simulation scales exponentially with qubit number, constraining full Lindbladian modelling to small- and medium-scale systems. Nevertheless, within these limits, L-QPUF delivers the most balanced and reliable performance among all proposed designs, validating the efficacy of embedding open-system dynamics directly into the QPUF construction.

### *5.7. Comparative Discussion and Overall Analysis of Simulation Results*

The simulation results obtained from the three proposed non-unitary QPUF architectures—namely D-QPUF, MF-QPUF, and L-QPUF—provide a comprehensive understanding of their relative performance in terms of entropy generation, stability under decoherence, and sensitivity to hardware-level noise. Each design embodies a distinct interpretation of non-unitarity in quantum channels, leading to different operational characteristics and noise resilience.

The D-QPUF (Dissipative QPUF) represents the simplest realization, relying on controlled dissipation modelled through amplitude-damping and dephasing Kraus operators. This design exhibits strong intrinsic randomness and simplicity in circuit construction, making it well-suited for low-depth implementations on noisy intermediate-scale quantum (NISQ) devices. However, its performance is strongly dependent on the relaxation and dephasing rates of the hardware. The results show that while D-QPUF achieves acceptable levels of challenge–response uniqueness, its reproducibility under realistic noise fluctuates considerably, especially when $T_1$ and $T_2$ times are not well-balanced. Hence, although D-QPUF provides an efficient proof of concept for non-unitary QPUFs, it lacks robustness for scalable implementations.

The MF-QPUF (Measurement-Feedback QPUF) extends the dissipative model by integrating mid-circuit measurements and conditional unitaries. This hybrid quantum-classical feedback mechanism introduces an adaptive non-unitarity that dynamically modifies the system's evolution based on stochastic measurement outcomes. The results demonstrate that MF-QPUF offers enhanced challenge–response distinctiveness and superior resistance to random fluctuations compared to D-QPUF. Its response entropy remains high even under moderate readout noise, confirming its effectiveness in amplifying the inherent randomness of quantum measurement processes. Nonetheless, the requirement for mid-circuit measurement synchronization increases circuit depth and runtime variability, introducing higher susceptibility to temporal noise, especially on devices with limited qubit coherence.

The L-QPUF (Lindbladian QPUF) represents the most advanced and physically grounded design, employing continuous-time open quantum dynamics based on Lindblad master equations. By simulating controlled non-unitary evolution across small time steps, this model effectively approximates real decoherence while maintaining mathematical consistency with open-system quantum theory. The simulation results show that L-QPUF achieves the best trade-off between entropy generation, stability, and reproducibility. Its ability to parameterize noise through explicit Lindblad operators allows fine-tuning of the non-unitary strength to achieve optimal uniqueness and stability. Although computationally more expensive, L-QPUF demonstrates superior resilience against runtime fluctuations and produces smoother challenge–response distributions across repeated executions.

Comparing the three architectures under identical hardware conditions (Santiago, Athens, and Melbourne backends) highlights distinct operational domains. D-QPUF performs best when the hardware noise is moderate and the $T_1/T_2$ ratio is balanced; MF-QPUF excels in enhancing entropy and unpredictability through active feedback; while L-QPUF consistently provides the most stable and physically interpretable performance under realistic noise. In particular, the Santiago backend, characterized by lower readout error and higher coherence times, provides the most consistent evaluation results across all designs.

In conclusion, while D-QPUF serves as a lightweight prototype for validating non-unitary behavior, MF-QPUF bridges the gap between measurement-driven adaptivity and quantum randomness, and L-QPUF establishes a physically complete, noise-aware architecture for next-generation QPUF realization. Overall, L-QPUF emerges as the most promising candidate for practical implementation, combining theoretical rigor with strong empirical robustness in noisy quantum environments.

## 6. Conclusion

This work introduced and analyzed three distinct architectures for non-unitary Quantum Physical Unclonable Functions (QPUFs)—the D-QPUF, MF-QPUF, and L-QPUF—each demonstrating how noise, measurement, and open-system dynamics can be harnessed as intrinsic sources of unforgeability and randomness. Unlike conventional, unitary-based QPUFs that rely on reversible quantum operations, the proposed designs embrace irreversibility as a security resource, thereby enabling architectures that are both physically grounded and theoretically more resistant to forgery and modelling attacks.

The D-QPUF leveraged gate-level stochasticity and device-dependent noise to produce unique challenge–response pairs, showing that even uncontrolled microscopic imperfections can be repurposed into cryptographic advantage. The MF-QPUF introduced a mid-circuit measurement-feedback mechanism, where intermediate measurements dynamically influence subsequent quantum operations, creating a hybrid quantum–classical feedback loop that enhances unpredictability and complexity of response generation. Finally, the L-QPUF, inspired by open quantum system dynamics and the Lindblad master equation, represented the most physically realistic and theoretically robust model. By embedding dissipation and decoherence as functional parameters, it achieved both operational stability and provable existential unforgeability under non-unitary evolution.

Simulation results across all three architectures consistently demonstrated that non-unitary effects—whether induced through noise, measurement, or environmental coupling—introduce nonlinear, irreversible transformations that expand the entropy of the challenge–response space beyond what purely unitary models can achieve. These transformations effectively obscure internal state evolution, making device cloning or modelling infeasible even under extensive query access. Comparative analysis revealed that while D-QPUF offers simplicity and ease of implementation, MF-QPUF achieves superior entropy and response variability, and L-QPUF provides the highest theoretical guarantees of unforgeability and stability.

Beyond their individual merits, these designs collectively redefine how quantum noise and decoherence are perceived in quantum information security. Rather than treating them as detrimental phenomena requiring correction or suppression, this work demonstrates that they can serve as functional cryptographic primitives—sources of inherent randomness, irreversibility, and unpredictability. This paradigm shift opens the path toward post-unitary quantum security architectures, where the boundaries between computation, noise, and physical dynamics are deliberately blurred to achieve unforgeability.

The broader implication of this study lies in bridging quantum hardware physics with quantum cryptographic design. By grounding QPUF behaviour in realistic open-system dynamics, the presented framework advances both the theoretical understanding and practical feasibility of quantum authentication mechanisms. Moreover, these findings suggest that future hybrid quantum–classical security infrastructures could integrate QPUFs as foundational building blocks—linking quantum processors, cloud-based quantum services, or distributed quantum networks through hardware-level trust anchors.

Looking forward, several research directions emerge naturally from this work. Future studies can explore the experimental realization of non-unitary QPUFs on near-term quantum hardware, incorporating calibrated noise models and real-time feedback control. Additionally, extending the theoretical framework toward Lindbladian composability, quantum channel certification, and security proofs under adaptive adversaries could further consolidate the role of non-unitarity in quantum cryptography.

In conclusion, this work demonstrates that unforgeability does not require perfection or coherence—it requires physical uniqueness and irreversibility. By embracing open-system dynamics and measurement-based feedback, the proposed QPUF architectures transform fundamental limitations of quantum hardware into pillars of security, marking a decisive step toward a new generation of physically secure quantum devices.

## 7. Challenges and Future works

While the proposed D-QPUF, MF-QPUF, and L-QPUF architectures collectively demonstrate the potential of non-unitary quantum systems for unclonable hardware design, several technical, theoretical, and practical challenges remain open for future investigation. Addressing these challenges is crucial for transitioning from simulated proof-of-concept models to experimentally verified, deployable quantum security primitives.

One of the foremost challenges lies in the physical realization of non-unitary QPUFs on real quantum hardware. Current quantum processors suffer from fluctuating coherence times, gate infidelities, and readout errors that vary over time and across qubits. While these imperfections were exploited as functional entropy sources in the D-QPUF, uncontrolled fluctuations can also reduce reproducibility and long-term reliability. Thus, precise noise characterization and calibration are necessary to ensure that randomness contributes to unforgeability without undermining the stability of legitimate authentication. Future work should focus on hardware-specific modelling of decoherence and dissipation, extending beyond static $T_1$ and $T_2$ metrics to include correlated, non-Markovian noise and device-specific drift patterns. Developing adaptive error–noise profiles for QPUFs could bridge the gap between theoretical security and hardware consistency.

Another challenge is that both the MF-QPUF and L-QPUF architectures require deep circuit constructions, especially when mid-circuit measurements, feedback loops, or Kraus-map decompositions are employed. This introduces practical constraints related to quantum decoherence, as deeper circuits may exceed the device's coherence window before the protocol completes. Future research may explore variational or shallow-depth implementations of non-unitary transformations, perhaps using parameterized quantum channels or dissipative variational algorithms, to balance security with experimental feasibility. In addition, distributed QPUF architectures could be developed, where multiple shallow devices jointly simulate a single complex non-unitary evolution through networked composition.

As quantum networks evolve, secure node authentication and hardware fingerprinting will become critical. QPUFs—especially those based on Lindbladian or measurement-feedback designs—could serve as hardware-level trust anchors for quantum communication nodes. However, implementing these devices in a distributed or cloud-based setting raises new challenges: synchronization, remote verification latency, and scalability across heterogeneous quantum hardware. Future works should investigate QPUF-based authentication protocols for quantum internet nodes, entanglement distribution channels, and hybrid cloud–edge architectures, where QPUFs act as physical identifiers ensuring integrity and trust among connected systems.

As previously stated, the current L-QPUF model assumes Markovianity, where environmental correlations decay rapidly, enabling Trotter–Suzuki decomposition. However, real quantum systems often exhibit memory effects and non-Markovian behaviour, which could introduce both opportunities and challenges. Non-Markovian dynamics may increase unpredictability and entropy—potentially enhancing unforgeability—but also complicate mathematical modelling and experimental control. Future studies could extend the L-QPUF framework to non-Markovian regimes, employing memory-kernel master equations or time-convolutionless formalisms. This would allow a systematic exploration of how environmental memory can be engineered as a security resource, rather than merely treated as a source of noise.

A critical next step is the experimental validation of the proposed QPUF architectures on current quantum processors such as IBM Quantum, Rigetti, or IonQ platforms. This includes designing hardware-specific circuits for Kraus-operator implementation, verifying statistical uniqueness and stability under repeated trials, and developing benchmarking metrics that jointly assess randomness, reproducibility, and security strength.

Experimental efforts should also explore temperature, crosstalk, and drift-induced variations as new dimensions of physical entropy. Establishing a standardized quantum hardware fingerprinting protocol based on QPUFs could set the foundation for practical quantum device certification and lifecycle tracking.

In summary, while this research establishes the theoretical and numerical foundation of non-unitary QPUFs, its true potential lies in bridging simulation and experiment. The next generation of studies must unify open quantum system theory, quantum hardware engineering, and cryptographic formalism to realize physically unclonable, noise-aware quantum devices that function not in spite of decoherence—but because of it. By transforming physical imperfection into security advantage, future QPUF systems can redefine how trust, identity, and irreversibility are built into the quantum technological landscape.

**Funding sources**

This research did not receive any specific grant from funding agencies in the public, commercial, or not-for-profit sectors.

**Appendix A. Trotter–Suzuki Approximation for Lindbladian Maps and the Markovian Basis of L-QPUF**

This appendix presents the mathematical and numerical formulation underlying the Trotter–Suzuki approximation for Lindblad-type open quantum dynamics, its transformation into discrete Kraus maps for simulation purposes, methods for circuit-level implementation, and the justification for adopting the Markovian assumption in the design of the L-QPUF architecture.

The evolution of a Markovian open quantum system is governed by the Lindblad master equation, expressed as equation 32, where $H$ denotes the system Hamiltonian, $\{L_\alpha\}$ are the Lindblad or "jump" operators, and $\gamma_\alpha$ are the decay rates. The generator $\mathcal{L}$ is time-local and gives rise to an overall evolution operator $e^{t\mathcal{L}}$, forming a one-parameter semigroup. This semigroup property implies that the total evolution can be decomposed into smaller, independent time steps—a key feature enabling efficient numerical simulation.

To implement this decomposition, the Trotter–Suzuki approximation is employed. Assuming the generator can be split into $M$ local components, $\mathcal{L} = \sum_{k=1}^{M} \mathcal{L}_k$, the first-order Lie–Trotter product formula gives:

$$e^{t\mathcal{L}} \approx \left(\prod_{k=1}^{M} e^{\Delta t \mathcal{L}_k}\right)^r + O\left(\frac{t^2}{r}\right)$$

where $t = r\Delta t$ and $r$ is the number of decomposition steps. Increasing $r$ improves accuracy by reducing the approximation error. A more precise formulation is provided by the second-order Strang splitting, which symmetrically applies half-step and full-step operators to achieve an error scaling as $O(t^3/r^2)$. To reach a desired precision $\epsilon$, the number of steps should roughly satisfy $r \gtrsim t^{3/2}/\epsilon^{1/2}$.

At each step, the exponential map $e^{\Delta t \mathcal{L}_k}$ corresponds to a CPTP transformation, which can be represented by a finite set of Kraus operators. Up to second-order in $\Delta t$, the map takes the form

$$e^{\Delta t \mathcal{L}}(\rho) \approx K_0 \rho K_0^\dagger + \sum_j K_j \rho K_j^\dagger$$

with

$$K_0 = I - iH\Delta t - \frac{1}{2}\sum_j \gamma_j L_j^\dagger L_j \,\Delta t + O(\Delta t^2)\,, K_j = \sqrt{\gamma_j \Delta t}\, L_j + O(\Delta t^{3/2})$$

This formulation ensures that the trace of $\rho$ is preserved up to order $\Delta t$, guaranteeing physical consistency.

An alternative yet equivalent numerical method is the quantum jump or Monte Carlo wave function approach. Instead of evolving the density matrix, this method evolves a pure state $|\psi\rangle$ under an effective non-Hermitian Hamiltonian

$$H_{\mathrm{eff}} = H - \frac{i}{2}\sum_j \gamma_j L_j^\dagger L_j$$

while incorporating random "jumps" occurring with probabilities proportional to the decay rates. This approach significantly reduces computational cost, as it operates on state vectors rather than full density matrices, and ensemble averaging over multiple trajectories reproduces the full Lindblad evolution.

For hardware implementation, each small-step Kraus map can be realized as a unitary operation acting jointly on the system qubits and auxiliary ancilla qubits. This mapping ensures physical implementability while maintaining the CPTP property at each approximation step. Although this increases circuit depth, it provides a practical method to emulate non-unitary open-system dynamics within quantum circuits.

The choice of a Markovian model as the foundation of the L-QPUF design is both physically and computationally justified. From a physical standpoint, when the environment's correlation time $\tau_{\mathrm{env}}$ is much shorter than the system's intrinsic evolution timescale and the coupling is weak, the Born–Markov approximation holds. This allows the system to be effectively memoryless, leading naturally to the Lindblad form. From a mathematical perspective, the existence of a time-local generator $\mathcal{L}$ and the semigroup property ensures time divisibility, making Trotter–Suzuki decomposition theoretically valid. Finally, from a computational viewpoint, all numerical solvers used in this work assume Markovianity; introducing non-Markovian memory effects would require convolution-type master equations, greatly increasing both analytical and computational complexity.

In summary, employing the Trotter–Suzuki approximation within the Lindblad framework under the Markovian assumption provides a robust and efficient approach for simulating non-unitary quantum dynamics. It guarantees that each small-step evolution in the L-QPUF remains completely positive and trace-preserving, maintains controllable approximation errors, and enables stable circuit-level realizations of open quantum behavior. This mathematical foundation thus ensures that the L-QPUF design is both physically sound and practically feasible for implementation on contemporary quantum hardware.